\documentstyle[11pt]{article}
\input amssymb.sty
\input amssymb.sty

\begin{document}
\newcommand{\Si}{\Sigma}
\newcommand{\tr}{{\rm tr}}
\newcommand{\ad}{{\rm ad}}
\newcommand{\Ad}{{\rm Ad}}
\newcommand{\ti}[1]{\tilde{#1}}
\newcommand{\om}{\omega}
\newcommand{\Om}{\Omega}
\newcommand{\de}{\delta}
\newcommand{\al}{\alpha}
\newcommand{\te}{\theta}
\newcommand{\vth}{\vartheta}
\newcommand{\be}{\beta}
\newcommand{\la}{\lambda}
\newcommand{\La}{\Lambda}
\newcommand{\D}{\Delta}
\newcommand{\ve}{\varepsilon}
\newcommand{\ep}{\epsilon}
\newcommand{\vf}{\varphi}
\newcommand{\G}{\Gamma}
\newcommand{\ka}{\kappa}
\newcommand{\ip}{\hat{\upsilon}}
\newcommand{\Ip}{\hat{\Upsilon}}
\newcommand{\ga}{\gamma}
\newcommand{\ze}{\zeta}
\newcommand{\si}{\sigma}
\def\bfa{{\bf a}}
\def\bfb{{\bf b}}
\def\bfc{{\bf c}}
\def\bfd{{\bf d}}
\def\bfe{{\bf e}}
\def\bfm{{\bf m}}
\def\bfn{{\bf n}}
\def\bfp{{\bf p}}
\def\bfu{{\bf u}}
\def\bfv{{\bf v}}
\def\bft{{\bf t}}
\def\bfx{{\bf x}}
\def\bfg{{\bf g}}
\def\bfS{{\bf S}}
\def\bfJ{{\bf J}}
\newcommand{\li}{\lim_{n\rightarrow \infty}}
\newcommand{\mat}[4]{\left(\begin{array}{cc}{#1}&{#2}\\{#3}&{#4}
\end{array}\right)}
\newcommand{\thmat}[9]{\left(
\begin{array}{ccc}{#1}&{#2}&{#3}\\{#4}&{#5}&{#6}\\
{#7}&{#8}&{#9}
\end{array}\right)}
\newcommand{\beq}[1]{\begin{equation}\label{#1}}
\newcommand{\eq}{\end{equation}}
\newcommand{\beqn}[1]{\begin{eqnarray}\label{#1}}
\newcommand{\eqn}{\end{eqnarray}}
\newcommand{\p}{\partial}
\newcommand{\di}{{\rm diag}}
\newcommand{\oh}{\frac{1}{2}}
\newcommand{\su}{{\bf su_2}}
\newcommand{\uo}{{\bf u_1}}
\newcommand{\SL}{{\rm SL}(2,{\mathbb C})}
\newcommand{\GLN}{{\rm GL}(N,{\mathbb C})}
\newcommand{\PGLN}{{\rm PGL}(N,{\mathbb C})}
\def\sln{{\rm sl}(N, {\mathbb C})}
\def\SLN{{\rm SL}(N, {\mathbb C})}
\newcommand{\gln}{{\rm gl}(N, {\mathbb C})}
\newcommand{\PSL}{{\rm PSL}_2( {\mathbb Z})}
\def\f1#1{\frac{1}{#1}}
\def\lb{\lfloor}
\def\rb{\rfloor}
\def\sn{{\rm sn}}
\def\cn{{\rm cn}}
\def\dn{{\rm dn}}
\newcommand{\ran}{\rangle}
\newcommand{\lan}{\langle}
\newcommand{\rar}{\rightarrow}
\newcommand{\upar}{\uparrow}
\newcommand{\sm}{\setminus}
\newcommand{\ms}{\mapsto}
\newcommand{\bp}{\bar{\partial}}
\newcommand{\bz}{\bar{z}}
\newcommand{\bA}{\bar{A}}
\newcommand{\bL}{\bar{L}}

\newcommand{\vtb}{\theta_{10}}
\newcommand{\vtc}{\theta_{00}}
\newcommand{\vtd}{\theta_{01}}

\def\frak{\mathfrak}
\def\gg{{\frak g}}
\def\gJ{{\frak J}}
\def\gH{{\frak H}}
\def\gS{{\frak S}}
\def\gL{{\frak L}}
\def\gG{{\frak G}}
\def\gk{{\frak k}}
\def\gK{{\frak K}}
\def\gl{{\frak l}}
\def\gh{{\frak h}}

\def\mC{{\mathbb C}}
\def\mZ{{\mathbb Z}}
\def\mR{{\mathbb R}}
\def\mN{{\mathbb N}}
\def\mL{{\mathbb L}}

\newcommand{\sect}[1]{\setcounter{equation}{0}\section{#1}}
\renewcommand{\theequation}{\thesection.\arabic{equation}}
\newtheorem{predl}{Proposition}[section]
\newtheorem{defi}{Definition}[section]
\newtheorem{rem}{Remark}[section]
\newtheorem{cor}{Corollary}[section]
\newtheorem{lem}{Lemma}[section]
\newtheorem{theor}{Theorem}[section]

\vspace{0.3in}
\begin{flushright}
 ITEP-TH-56/01\\
\end{flushright}
\vspace{10mm}
\begin{center}
{\Large{\bf Hitchin Systems - Symplectic Hecke Correspondence and
Two-dimensional Version}
}\\
\vspace{5mm}
\today\\
\vspace{5mm}
A.M.Levin\footnote{ On leave from Institute of Oceanology, Moscow, Russia} \\
{\sf Max Planck Institute of Mathematics, Bonn, Germany,} \\
{\em e-mail alevin@wave.sio.rssi.ru}\\
M.A.Olshanetsky\footnote{ On leave from
Institute of Theoretical and Experimental Physics, Moscow, Russia}
\\
{\sf Max Planck Institute of Mathematics, Bonn, Germany,}\\
{\em e-mail olshanet@gate.itep.ru}\\
A.Zotov \\
{\sf Institute of Theoretical and Experimental Physics, Moscow, Russia,}\\
{\em e-mail zotov@gate.itep.ru}\\
\vspace{5mm}
\end{center}
\begin{abstract}
The aim of this paper is two-fold. First,
we define symplectic maps between Hitchin systems related to holomorphic
bundles of different degrees. We call these maps the Symplectic Hecke
Correspondence (SHC) of the corresponding Higgs bundles.
They are constructed by
means of the Hecke correspondence of the underlying holomorphic bundles.
SHC allows to construct B\"{a}cklund transformations in the
Hitchin systems defined over Riemann curves with  marked points.
We apply the general scheme to the elliptic Calogero-Moser (CM) system and
construct SHC to an integrable $\SLN$ Euler-Arnold top
(the elliptic $\SLN$-rotator).
Next, we propose a generalization of the Hitchin approach to 2d integrable
theories related to the Higgs bundles of infinite rank. The main example
is an integrable two-dimensional version of the two-body elliptic  CM system.
The previous construction allows to define SHC between
 the  two-dimensional elliptic CM system and the Landau-Lifshitz equation.
\end{abstract}

\tableofcontents

\section{Introduction}
\setcounter{equation}{0}

Nowadays many examples of integrable one-dimensional and
two-dimensional models are known. The problem of listing all
of them, up to some equivalence, was solved for some particular forms
of two-dimensional models \cite{MS}. The recently developed
conception of duality for one-dimensional models
\cite{FGNR} can shed light on the classification problem in analogy with
 string theory. In spite of this progress
we are still far from understanding the structure of this universe.
Therefore, the classification of
integrable systems, apart from solving any individual equation, continues to be
an actual task.
We will consider  integrable systems that have the Lax or
Zakharov-Shabat representations. In these cases the gauge transformations
of the accompanying linear equations lead essentially to the same systems,
though their equations of motion
differ in a significant way. For example, the non-linear
Schr\"{o}dinger model is gauge equivalent to the isotropic Heisenberg
magnetic \cite{FT2}.
In such a manner the integrable system should be classified up to
gauge equivalence, though it is not the only equivalence principle
in their possible classifications.
The crucial and delicate point of this approach is the exact definition
of allowed gauge transformations, and it will be discussed here.

We restrict ourselves to Hitchin systems
 \cite{H} and their two-dimensional generalizations that we will construct.
 The Hitchin construction
establishes relations between finite dimensional integrable systems and
the moduli space of holomorphic vector bundles over Riemann curves.
The phase space of the integrable system is the cotangent bundle to the
moduli space and the dual variables $\Phi$ are called the Higgs fields.
The pair $(E,\Phi)$, where $E$ is a holomorphic bundle, is called the
Higgs bundle.
The Lax representation arises immediately in this scheme as the equation of
motion and the Lax operator is just the Higgs field defined on shell.
The $C^\infty$ gauge transformations of the Lax pair define the equivalent
 holomorphic bundles. The different gauge fixing conditions give equivalent
integrable systems.

We consider the generalization of the Hitchin systems based on the
quasi-parabolic Higgs bundles \cite{Si},
where the Higgs fields are allowed to have the
first order poles at the marked points
 on the base curve.
 The gauge transformations preserve the flag structures that arise at
the marked points.
 The corresponding integrable systems were considered
in \cite{Bea,Ma,Ne,ER1}.
We loosen the smoothness condition
of the gauge transformations and allow them to have a simple zero or a pole
at one of the marked points.  This type of  gauge
transformations (the upper and lower symplectic Hecke correspondence (SHC))
 is suggested by the geometric Langlands program. SHC changes the degree
of the underlying bundles on $\pm 1$.
We assume, that HC is consistent with flag structures on the source and
target bundles. It allows to choose a canonical form of the modifications.
HC can be lifted as the symplectic correspondence (SHC) to the Higgs bundles.
In this way SHC define a map of Hitchin systems related to bundles of
different degrees.
One can consider an arbitrary chain of consecutive SHC
attributed to different marked points.
If the resulting transformation preserves the degree of bundle, then
it defines the B\"{a}cklund transformations of the Hitchin system related to
the initial bundle, or the integrable discrete time map \cite{Ve}.
Our construction is similar
to the scheme proposed by Arinkin and Lysenko \cite{AL} in the investigations
of the flat $\SL$- bundles over rational curves and the geometric structure of the
B\"{a}cklund transformations in  the Painl\'{e}ve 6 system \cite{Ok}.

As an example,
we consider a trivial holomorphic $\SLN$-bundle $E^{CM}$ (deg$(E^{CM})=0$)
 over
an elliptic curve with a marked point. The corresponding quasi-parabolic
Higgs bundle leads
 to the elliptic $N$-body Calogero-Moser system (CM system).
The upper SHC defines a map of the Higgs bundle related to $E^{CM}$ to the
Higgs bundle $(E^{rot},\Phi^{rot})$ with deg$(E^{rot})=1$.
SHC is generated by the
$N$-th order matrix $\Xi$
with theta-functions depending on coordinates of the particles as the matrix
elements.
The system $(E^{rot},\Phi^{rot})$
is the integrable $\SLN$-Euler-Arnold top ($\SLN$-elliptic rotator).
The Lax pair  for this top was proposed earlier \cite{STSR}.
The consecutive upper and
lower SHC define the B\"{a}cklund transformations of the both
systems.
A construction of this type was suggested in \cite{Vak} for
studding the B\"{a}cklund transformations of the Ruijsenaars model.
Another way to find a B\"{a}cklund transformation
is achieved by applying $N$ consecutive upper modifications,
since they lead to equivalent Higgs bundles.

In the second part of the paper we try to gain insight into interrelation
between integrable theories in dimension one and two.
It is known that some one-dimensional integrable systems can be extended
to the two-dimensional case without sacrificing the integrability.
For example, the Toda field theory comes from the corresponding Toda
lattice. To understand this connection
we apply the Hitchin construction to two-dimensional systems.
For this purpose
we consider infinite rank bundles over the Riemann curves with marked points.
The transition group of the bundles is the central extended loop group
$\hat{L}(\GLN)$. If the central charge vanishes the theory in essence
becomes one-dimensional.
In the two-dimensional situation  the Higgs field is a
 $\gln$ connection on a circle $S^1$.
In addition, we put coadjoint orbits of $\hat{L}(\GLN)$ at the marked
points and in this way introduce the quasi-parabolic structure
on the Higgs bundle of infinite rank. The monodromy
of the Higgs field is a generating function for the infinite number of
conservation laws.
The equations of motion on the reduced phase space are the Zakharov-Shabat
equations.
The similar class of the Hitchin type systems from a different point of
view was introduced recently by Krichever \cite{K}.
We consider in detail the case of $\hat{L}(\SL)$-bundle over an elliptic
curve with $n$ marked points. The Higgs bundle corresponds to the
two-dimensional version of the elliptic Gaudin system. For the 1
marked point case
we come to the 2d two-body elliptic CM theory.
The upper SHC is working in the two-dimensional
situation as well. It leads to the map of the
2-body elliptic CM field theory to
the Landau-Lifshitz equation.\footnote
{The equivalence of these models was pointed out by A.Shabat.}
To summarize we consider here the following diagram:

\bigskip
$$
\begin{array}{ccc}
\fbox{$
\begin{array}{ccc}
2-{\rm body}&{\rm elliptic}&CM\\
 &{\rm system}&
\end{array}
$}&
\longrightarrow &
\fbox{$
\SL-{\rm elliptic ~rotator}
$}\\
\downarrow&    &\downarrow\\
\fbox{$
\begin{array}{ccc}
2-{\rm body}&{\rm elliptic}&CM\\
 &{\rm field~theory}&
\end{array}
$}&
\longrightarrow &
\fbox{Landau-Lifshitz equation}
\end{array}
$$
\begin{center}
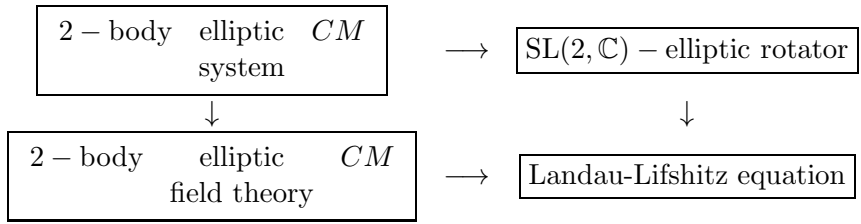
Figure 1: Interrelation in integrable theories
\end{center}

\bigskip
In fact, the upper SHC can be applied to the $\SLN$ case.
The quadratic Hamiltonian
of the $N$-body elliptic CM field theory was constructed in \cite{K},
but the $\SLN$ generalization of the Landau-Lifshitz equation is unknown.

It should be mentioned that the quantum version of $\SLN$ SHC
 has appeared in a different context long ago \cite{FT}.
It was defined as a twist transformation of the quantum $R$-matrices, and
Hasegawa \cite{Has} has constructed such types of twists that transform
the dynamical elliptic $R$-matrix of Felder \cite{Fe}
 to the non-dynamical $R$-matrix of Belavin \cite{Be}.
It was proved \cite{Ar} that the dynamical $R$-matrix corresponds to the
elliptic Ruijsenaars system  \cite{R}. The later
 is the relativistic deformation of the elliptic CM system. In this way
the Hasegawa twist is the quantization of SHC we have constructed,
since the elliptic CM system and the elliptic Ruijsenaars system are governed
by the same $R$-matrix \cite{S}.

\section{Hitchin systems in the  \u{C}ech description}
\setcounter{equation}{0}

In this section we consider vector bundles with structure group $G=\GLN$,
or any simple complex Lie group.

\subsection{The moduli space of holomorphic quasi-parabolic bundles
 in the  \u{C}ech description.}
Let $E$ be a trivial rank $r$ holomorphic vector bundle over a
Riemann curve $\Si_n$ with $n$ marked points. Consider a covering of $\Si_{n}$
 by open disks ${\cal U}_a,~a=1,2\ldots$.
Some of them may contain one marked point $w_\al$. The holomorphic structure
on $E$ can be described by
the  differential $d^{''}$. On ${\mathcal U}_a$ it can be represented as
$$
d^{''}=\bp_a+\bA_a,~\bA_a=h_{a}^{-1}\bp_a h_{a}, ~~\bp_a=\frac{\p}{\p\bz_a}
$$
 where $z_a$ is a local coordinate on ${\mathcal U}_a$ , and $h_a$ is a
$C^\infty$ $G$-valued
function on ${\mathcal U}_a$.  It is a  section of the local sheaf
$\Om_{C^\infty}^0(\Si_n,{\rm Aut}~E)$.

The  transition functions $g_{ab}=h_ah_b^{-1}$ are defined on the intersections
${\cal U}_{ab}={\cal U}_a\cap{\cal U}_b$. They are holomorphic since
$\bA_a=\bA_b$ on ${\cal U}_{ab}$
$$
g_{ab}\in \Om_{hol}^0({\cal U}_{ab},{\rm Aut}~ E)\,.
$$

The transformation $h_{a}\to f_{a}h_{a}$ by a function holomorphic
on ${\mathcal U}_a$
($f_a\in \Om_{hol}^0({\cal U}_{a},{\rm Aut}~ E)  $) does
not change $\bA_a$. Similarly, the transformation $h_{b}\to f_{b}h_{b}$
by  $f_b\in \Om_{hol}^0({\cal U}_{b},{\rm Aut}~ E)$ does
not change $\bA_b$.
Then the holomorphic structures described by the transition
functions $g_{ab}$ and $f_{a}g_{ab}f_{b}^{-1}$ are equivalent.
 Globally we have
the collection of  transition maps
\beq{a16}
{\cal L}_{\Si}^C=\{g_{ab}(z_a)=h_a(z_a)h^{-1}_b(z_b(z_a)),~z_a\in
{\cal U}_{ab},~a,b=1,2\ldots ,\}\,.
\eq
They define  holomorphic structures on $E$ or $P={\rm Aut}E$ depending
on the choice of the representations.

The definition of the holomorphic structures
by the transition functions works as well in the case if ${\rm deg}(E)\neq 0$
($G=\GLN$).
They should satisfy the cocycle condition
\beq{coc}
g_{ab}(z)g_{bc}(z)g_{ca}(z)={\rm Id}, ~~
z\in {\cal U}_{a}\cap{\cal U}_{b}\cap{\cal U}_{c}\,,
\eq
and
\beq{a16a}
g_{ab}=g_{ba}^{-1}.
\eq
The degree of the bundle $E$ is defined as the degree of the linear bundle
$L=\det g$.

We choose an open subset of stable holomorphic
structures ${\cal L}_{\Si}^{C,st}$ in ${\cal L}_{\Si}^C$.
The gauge group ${\cal G}_{\Si}^{hol}$ acts as the automorphisms of
 ${\cal L}_{\Si}^{C,st}$
\beq{a17}
g_{ab}\rar f_ag_{ab}f^{-1}_b,~f_a=f(z_a),~
f_b=f_b(z_b(z_a)),~
f\in{\cal G}_{\Si}^{hol}\,.
\eq
We prescribe the local behavior of the gauge transformations
${\cal G}_{\Si}^{hol}$  at the marked points.
Let
$$
P_1,\ldots,P_\al,\ldots,P_n
$$
be parabolic subgroups of $G$ attributed to the marked points.
Then we assume that
\beq{gg}
f_a=\left\{
\begin{array}{ll}
\ti{f}^{(0)}_\al+z_\al f_\al^{(1)}+\ldots, &
\ti{f}^{(0)}_\al\in P_\al,~{\rm if}~z_\al=z-w_\al,~w_\al~
{\rm is~ a~marked~point},
\\
f_a^{(0)}+z_af_a^{(1)}+\ldots,  & f_a^{(0)}\in G~{\rm if}~a\neq\al,~
({\cal U}_a~{\rm does ~not~contain~a~marked~point})\,.
\\
\end{array}
\right.
\eq

It follows from (\ref{a17}) that the left action of the gauge group
 at the marked points preserves the flags
\beq{fl}
E_\al(s)\sim P_\al\setminus G,~~
E_\al=Fl_1(\al)\supset\cdots\supset Fl_{s_\al}(\al)\supset Fl_{s_\al+1}(\al)=0\,.
\eq
The moduli space of the stable holomorphic bundles
 ${\cal M}_n(\Si,G)$ with the quasi-parabolic structure at the marked points
is defined in Ref.\cite{MeS} as the factor space under this action
\beq{a18}
{\cal M}_n={\cal G}_{\Si}^{hol}\backslash{\cal L}_{\Si}^{C,st}\,.
\eq
For $G=\GLN$ we have a disjoint union of
components labeled by the corresponding degrees $d=c_1(\det E)$ :
\mbox{${\cal M}_n(\Si,G)=\bigsqcup{\cal M}^{(d)}_n$.}

The tangent space to  ${\cal M}_n(\Si,G)$ is isomorphic to
$h^1(\Si,{\rm End}E)$. Its dimension can be extracted from
 the Riemann-Roch theorem and for curves without marked points $(n=0)$
$$
\dim h^0(\Si,{\rm End}E)-\dim h^1(\Si,{\rm End}E)=
(1-g)\dim G\,.
$$
For stable bundles $h^0(\Si,{\rm End}E)=1$ and
$$
\dim{\cal M}_0(\Si,G)=(g-1)N^2+1
$$
for $\GLN$, and
$$
\dim{\cal M}_0(\Si,G)=(g-1)\dim G
$$
for simple groups.
For elliptic curves one has
$$
\dim h^1(\Si,{\rm End}E)=\dim h^0(\Si,{\rm End}E),
$$
and
\beq{2.0}
\dim{\cal M}_0^d= {\rm g.c.d.} (N,d)\,.
\eq
In this case the structure of the moduli space  for the trivial bundles
(i.e. with ${\rm deg}(E)=0$ and, for example, for bundles with
${\rm deg}(E)=1$ are different. We use this fact below.

For the quasi-parabolic bundles we have
\beq{pb}
\dim{\cal M}_n^d=\dim{\cal M}_0^d+
\sum_{\al=1}^nf_\al\,,
\eq
where $f_\al$ is the dimension of the flag variety $E_\al$.
In particular, for $G=\GLN$, we get
\beq{pb1}
f_\al=\oh\left(
N^2-\sum_{i=1}^{s_\al}m_i^2(\al)
\right),~
m_i(\al)=\dim Fl_i(\al)-\dim Fl_{i+1}(\al)\,.
\eq

The space ${\cal L}_{\Si}^C$ is a sort of a lattice 2d gauge theory.
Consider the skeleton
 of the covering $\{{\cal U}_a,~a=1,\ldots\}$. It is an oriented graph
whose vertices $V_a$ are some fixed inner points in ${\cal U}_a$ and
edges $L_{ab}$
connect those $V_a$ and $V_b$ for whose $U_{ab}\neq\emptyset$.
We choose an orientation of the graph, saying that $a>b$ on the edge
 $L_{ab}$ and put
the holomorphic function $z_b(z_a)$ which defines the holomorphic map
from ${\cal U}_a$ to ${\cal U}_b$.
Then the space ${\cal L}_{\Si}^C$ can
be defined by the following data. To each edge $L_{ab},~a>b$
 we attach a matrix
valued function $g_{ab}\in G$ along with $z_b(z_a)$.
The gauge fields $f_a$ are living on the
vertices $V_a$ and the gauge transformation is given by (\ref{a17}).

\subsection{Hitchin systems.}
The Hitchin systems in the \u{C}ech description can be constructed in
the following way \cite{LO}.
We start from the cotangent bundle
$T^*{\cal L}_{\Si_n}^C$ to the holomorphic structures on
$P={\rm Aut}E$  (\ref{a16}). Now
\beq{bb4}
T^*{\cal L}_{\Si_n}^C=\{\eta_{ab},g_{ab}|
~\eta_{ab}\in\Om_{hol}^{(1,0)}({\cal U}_{ab},({\rm End}E)^*),~
g_{ab}\in\Om_{hol}^0({\cal U}_{ab},P)\}\,.
\eq
 The one forms $\eta_{ab}$ are called the Higgs fields.
This bundle can be endowed with a symplectic structure by means of
the Cartan-Maurer one-forms on $\Om_{hol}^0({\cal U}_{ab},P)$.

Let $\G_a^b(\be\ga)$ be an oriented edge in ${\cal U}_{ab}$
 with the end points
in the triple intersections
$\be\in {\cal U}_{abc}={\cal U}_{a}\cap{\cal U}_{b}\cap{\cal U}_{c}$,
$\ga\in {\cal U}_{abd}$. The fields $\eta_{ab},g_{ab}$ are attributed
 to the edge $\G_a^b(\be\ga)$. If we change the orientation
$\G_a^b(\be\ga)\to\G_b^a(\ga\be)$ the fields should be replaced on
$g_{ba}=g^{-1}_{ab}$ (see (\ref{a16a})) and
\beq{b4a}
\eta_{ab}(z_a)=g_{ab}(z_a)\eta_{ba}(z_b(z_a))g_{ab}^{-1}(z_a)\,.
\eq
For this reason the integral
\beq{f}
\int_{\G_a^b(\be\ga)}\tr\left(\eta_{ab}(z_a)Dg_{ab}g_{ab}^{-1}(z_a)\right)
\eq
is independent on the orientation.

We can put the data (\ref{bb4}) on the  graph $\{\G_a^b\}$ corresponding to the
covering $\{{\cal U}_a\}$. Taking into account (\ref{f}) we define
the symplectic structure
\beq{bb5}
\om^C=\sum_{\rm edges}
\int_{\G_a^b(\be\ga)}D\tr\left(\eta_{ab}(z_a)Dg_{ab}g_{ab}^{-1}(z_a)\right)\,.
\eq
Since $\eta_{ab}$ and $g_{ab}$ are both holomorphic in ${\cal U}_{ab}$,
the integral is
independent on the choice of the path $\G_a^b$ within ${\cal U}_{ab}$.
It is worthwhile to note that the cocycle condition (\ref{coc}) does not
yield the additional constraints.

The symplectic form is invariant under the gauge transformations (\ref{a17})
 supplemented by
\beq{bb6}
\eta_{ab}\rar f_a\eta_{ab}f_a^{-1}.
\eq
The set of invariant
commuting Hamiltonians on $T^*{\cal L}_{\Si}^C$ is
\beq{bb7}
I^C_{j,k}=\sum_{\rm edges}\int_{\G_a^b(\be\ga)}
\nu^C_{(j,k)}(z_a)\tr(\eta_{ab}^{d_j}(z_a))\,, ~~(k=1,\ldots,n_j)
\eq
where $d_j$ are the orders of the
basic invariant polynomials corresponding to $G$ and
$\nu^C_{j,k}$ are $(1-d_j,0)$-differentials. They are related locally
to the $(1-j,1)$-differentials by $\nu^D_{j,k}=\bp \nu^C_{j,k}$
and
$$
n_j=h^1(\Si,{\cal T}^{\otimes (d_j-1)})=
(2d_j-1)(g-1)+(d_j-1)n,~~ (j=1,\ldots,r)\,.
$$
for the simple groups, and
$$
n_j=\left\{
\begin{array}{ll}
(2j-1)(g-1)+(j-1)n, & (j=2,\ldots,N)\\
g, &j=1
\\
\end{array}
\right.
$$
for $\GLN$.
The total number of independent Hamiltonians is equal to
$$
\sum_{j=1}^Nn_j={\cal M}^d_0+\oh r(r+1)n\,.
$$
This number is greater than the dimension of the moduli
space ${\cal M}^d_n$ (\ref{pb}). There are $rn$ highest weight integrals,
($j=r$), that become Casimir elements of coadjoint orbits
after the symplectic
reduction, that we will  consider below.

Perform the symplectic reduction with respect to the gauge action
(\ref{a17}), (\ref{bb6}) of ${\cal G}_{\Si_n}^{hol}$ (\ref{gg}).
The moment map is
$$
\mu_{{\cal G}_{\Si}^{hol}}(\eta_{ab},g_{ab}):~T^*{\cal L}_{\Si}^C\rar
{\rm Lie}^*({\cal G}_{\Si}^{hol}).
$$
Here the Lie coalgebra ${\rm Lie}^*({\cal G}_{\Si}^{hol})$
is defined with respect to the pairing
$$
\sum_{\rm edges}\int_{\G_a^b(\be\ga)}\tr(\xi_a\ep_a),~~
\ep_a\in{\rm Lie}({\cal G}_{\Si}^{hol}).
$$
Then locally we have
\beq{co}
\xi_a=\left\{
\begin{array}{ll}
\left(
z^{-1}_a\ti{\xi}_a+z^{-2}_a\xi_a^{(-2)}+\ldots
\right)dz_a, &
\ti{\xi}_a\in{\rm Lie}^*(P_\al),~
({\cal U}_a~{\rm contains~ a~marked~point}~w_\al)\\
\left(z^{-1}_a\xi_a^{(-1)}+z^{-2}_a\xi_a^{(-2)}+\ldots
\right)dz_a,  &\xi_a^{(-1)}\in{\rm Lie}^*(G)
({\cal U}_a~{\rm does ~not~contain~w_\al})\,.
\\
\end{array}
\right.
\eq
The canonical gauge transformations (\ref{a17}),(\ref{bb6}) of the
symplectic form (\ref{bb5}) are generated by
the Hamiltonian
$$
F_{\ep^{hol}}=
\sum_{\rm edges}
\int_{\G_a^b(\be\ga)}
\tr(\eta_{ab}(z_a)\ep_a^{hol}(z_a))-
\tr(\eta_{ab}(z_a)g_{ab}(z_a)\ep_b^{hol}(z_b(z_a))g_{ab}(z_a)^{-1})=
$$
$$
=\sum_a\int_{\G_a}\sum_b\tr(\eta_{ab}(z_a)\ep_a^{hol}(z_a))\,,
$$
where $\G_a$ is an oriented contour around ${\cal U}_a$.

 The non-zero moment
is fixed in a special way at the neighborhoods of the marked points.
Let $\ti{G}_\al\subset P_\al$ be the maximal semi-simple subgroup of the
parabolic group $P_\al$ defined at the marked point $w_\al$.
We drop for a moment the index $\al$ for simplicity.
We choose an ordering in the Cartan subalgebra $\gh\in$Lie$(G)$, which
is consistent with the embedding $P\subset G$. Let
$\ti{\gh}=\gh\cap\ti{G}$ be the Cartan subalgebra in $\ti{G}$.
Consider the orthogonal  decomposition of $\gh^*$
$$
\gh^*=\ti{\gh}^*+\gh'^*\,.
$$
We fix a vector $p^{(0)}\in \gh^*$ such that it is a generic element in
$\gh'^*$ and
\beq{orb}
\lan  p^{(0)},\ti{\gh}^*\ran=0\,,
\eq
where $\lan~,~\ran$ is the Killing scalar product in $\gh^*$.
Since  $\gh'^*\subset$Lie$^*(P)$, we can take
 $\mu_{{\cal G}_{\Si}^{hol}}$ in the form
\beq{mm}
\mu_{{\cal G}_{\Si}^{hol}}=\mu_0=\sum_{\al=1}^np_\al^{(0)}z_\al^{-1}dz_\al,
~ p^{(0)}\in\gh'^*    \,,
\eq
where $z_\al=z-w_\al$ is a local coordinate in ${\cal U}_\al$.
The moment equation $\mu_{{\cal G}_{\Si}^{hol}}=\mu_0$ can be read off
from $F_{\ep^{hol}}$. It follows from the definition of
${\rm Lie}^*({\cal G}_{\Si}^{0,hol})$ that $\eta_{ab}$ is the boundary value
of some
holomorphic or meromorphic one-form $H_a$ defined on ${\cal U}_a$ via
\beq{b8a}
\eta_{ab}(z_a)=H_a(z_a),~{\rm for}~z_a\in{\cal U}_{ab},~H_a\in
\Om^{(1,0)}_{hol}({\cal U}_a,{\rm End}^*(E))\,,
\eq
where
\beq{pol}
H_a=\left\{
\begin{array}{ll}
z_a^{-1}p^{(0)}_\al+H_a^{(0)}+
z_a H_a^{(1)}+\ldots, & {\rm if}~{\cal U}_a~{\rm
contains~ a~marked~point}~w_\al\\
H_a^{(0)}+z_aH_a^{(1)}+\ldots,  & {\rm if}~{\cal U}_a~{\rm does~ not~
contain~a~marked~point\,.}
\\
\end{array}
\right.
\eq
 The gauge fixing means that the transition functions
$g_{ab}$ are elements of the moduli space ${\cal M}^d_n(\Si,E) $. The symplectic quotient
\beq{hb}
{\cal H}_n^d={\cal G}_\Si^{hol}\backslash\backslash T^*{\cal L}_{\Si}^C=
{\cal G}_\Si^{hol}\backslash\mu^{-1}(\mu_0)
\eq
is called the Higgs bundle with the quasi-parabolic structures.
We set off the zero modes $g^{(0)}_{\al b}$
of  the transition functions
in the symplectic form on the reduced space  (see (\ref{bb5}))
\beq{sf}
\om^C=\sum_{\rm edges}
\int_{\G_a^b(\be\ga)}D\tr\left(\eta_{ab}(z_a)Dg_{ab}g_{ab}^{-1}(z_a)\right)+
2\pi i\sum_{\al=1}^n\sum_bD
\tr\left(p^{(0)}_\al Dg^{(0)}_{\al b}(g^{(0)}_{\al b})^{-1}\right)\,.
\eq
 The last
sum defines the Kirillov-Kostant symplectic forms
on the set of coadjoint orbits\\
 \mbox{
${\cal O}(n)=({\cal O}_1,\ldots{\cal O}_\al,\ldots,{\cal O}_n)$,}
where
\beq{tor}
{\cal O}_\al=\{p_\al\in {\rm Lie}^*(G)~|~p_\al=
(g^{(0)}_\al)^{-1} p^{(0)}_\al g^{(0)}_\al\}\,.
\eq
Note that $\dim({\cal O}_\al)=2f_\al$ (\ref{pb1}).

\begin{rem}
It is possible to construct another type of orbit ${\cal O}'_\al$
of the same dimension.
There exist elements  $p_\al^{'(0)}$ that belong to the complements of
Lie$^*(\ti{G}_\al)$ in Lie$^*(P_\al)$ such that
 the orbit
$$
{\cal O}'_\al=\{p_\al=
(g^{(0)}_\al)^{-1} p^{'(0)}_\al g^{(0)}_\al\}\,.
$$
 is symplectomorphic to the cotangent bundles  to the corresponding flags
 $E_\al$ (\ref{fl}) without the zero section
$T^*E_\al\setminus O(E_\alpha)$,  while ${\cal O}_\al$ (\ref{tor})
is a torsor over ${\cal O}'_\al$. Globally, ${\cal H}^d_n$ (\ref{hb})
is a torsor over $T^*{\cal M}^d_n$.
\end{rem}

\subsection{Standard description of the Hitchin system.}

The standard approach of the Hitchin systems \cite{H}
is based on the description
of the holomorphic bundles in terms of the operator $d''$.
The upstairs phase space has the form
\beq{ab1}
T^*{\cal L}_{\Si_n}^D=
\{\Phi,d''~|~\Phi\in \Om^{(1,0)}_{C^{\infty}}(\Si_n,{\rm End}^*~E)\}\,,
\eq
where $\Phi$ is called the Higgs field. The symplectic form
\beq{ab2}
\om^D=\int_{\Si_n}\tr(D\Phi\wedge D\bA)
\eq
 is invariant under the action of the gauge group
$$
{\cal G}_\Si^{C^\infty}=\{f\in\Om_{C^\infty}^0(\Si_n, {\rm Aut}~V)\},
$$
\beq{ab3}
\Phi\to f^{-1}\Phi f\,,~~\bA\to f^{-1}\bp f+ f^{-1}\bA f\,.
\eq
The gauge invariant integrals take the evident form (compare with (\ref{bb7}))
\beq{ab4}
I^D_{j,k}=\int_{\Si_n}
\nu^D_{(j,k)}\tr(\Phi^{d_j})\,, ~~(k=1,\ldots,n_j)
\eq
where $\nu^D_{j,k}$ are  $(1-j,1)$-differentials on $\Si_n$.
The symplectic reduction with respect to this action leads to the moment
map
$$
\mu~:~T^*{\cal L}_{\Si_n}^D\to {\rm Lie}^*({\cal G}_\Si^{C^\infty})~~
\mu=\bp\Phi+[\bA,\Phi]\,.
$$
The Higgs field $\Phi$ is related to $\eta$ in a simple way
$$
\eta_{ab} =h_{a}^{-1}\Phi h_{a}|_{{\cal U}_{ab}}\,,
$$
and $\bA_a=h_{a}^{-1}\bp_a h_{a}$.
The holomorpheity of $\eta$ is equivalent to the equation $\mu(\Phi,\bA)=0$,
 and $\Phi$ has the same simple poles as $H_a$ (\ref{b8a}).
 For simplicity, we call
$\eta$ the Higgs field.
The bundle $E$ equipped with the one-form $\eta$ is
called the Higgs bundle.

\subsection{ Modified \u{C}ech description of the moduli space.}
We modify the \u{C}ech description of
 the moduli space of $\GLN$-vector bundles in the following way.
Consider a formal (or rather small) disk $D$
embedded into $\Sigma$ in such way that
its center maps to the point $w$.

Consider first the case of $G=\PGLN$-bundles.
The moduli
space  ${\cal M}_n^d$ is  the quotient of the space ${\cal G}_{D^*}$ of
$G$-valued functions $g$ on the punctured disk $D^*$
by the right action of the group
${\cal G}_{out}$
of  $G$-valued holomorphic functions
on the complement to $w$ and by the left
action of the group ${\cal G}_{int}$
of $G$-valued holomorphic functions
on the disk:
$$
{\cal M}_n^d= {\cal G}_{int}\backslash
{\cal G}_{D^*}/{\cal G}_{out},\quad
g\to h_{int}gh_{out}\,.
$$
We assume that these transformations preserve the quasi-parabolic structure
of the vector bundle $E$.

Now consider $\GLN$-bundles. The group $\GLN$ is not
semi-simple. One has an action of the Jacobian
$Jac(\Si)$ on the moduli space of vector
bundles by the tensor multiplication, and the quotient is equal to
the space of $\PGLN$-bundles.
This follows from the exact sequence
$$
1\to{\cal O}^*\to {\rm GL}(N,{\cal O})\to{\rm PGL}(N,{\cal O})
\to 1\,.
$$
Hence locally
the moduli space of vector bundles
is product of the Jacobian of the curve
and moduli space of $\PGLN$-bundles.
We associate to the pair $(g,L)$
the bundle which is equal to $\mC^N\otimes L$ on the
 complement of a point, and the transition function
on the punctured disk is $g$.

Assume for simplicity that there is only one marked point and it coincides
with the center of $D^*$.
Let $z$ be the local coordinate on $D^*$. Then the gauge group
${\cal G}_{D^*}$ can be identify with the loop group $L(\GLN)$.
A parabolic subgroup of $L(\GLN)$
has the form
$$
{\cal G}_{int}\sim P\cdot\exp L^+(\gln),~~
L^+(\gln)=\sum_{j>0}\gg_jz^j, \gg_j\in\gln\,,
$$
where $P$ is a parabolic subgroup in $\GLN$ . The quotient
$LF(s)={\cal G}_{int}\backslash{\cal G}_{D^*}$ is the infinite-dimensional
flag variety, corresponding to the finite-dimensional flag
$E(s)$ (see (\ref{fl}))
\beq{fl1}
LFl(s)=\cdots\supset LFl_{r,k} \supset LFl_{r+1,k}\supset\cdots \supset LFl_{s,k}
\supset LF_{0,k-1}\supset\cdots\,,
\eq
$$
LFl_{r,k}=z^k Fl_r+\sum_{j< k}Ez^j)\, ~~(LFl_{s+1,k}=LFl_{0,k-1})\, .
$$

The $\GLN$ Higgs bundle ${\cal H}_n^d$ (\ref{hb}) can be
identified with the Hamiltonian quotient
$$
{\cal G}_{in}\backslash\backslash
T^*{\cal G}_{D^*}\times T^*Jac(\Sigma)//{\cal G}_{out}\,.
$$
 The cotangent bundle of
$T^*{\cal G}_{D^*}$ is identified with the space of pairs
$(g,\eta)$, where $\eta$ is a Lie$^*(G)$-valued one-form.
 The canonical one-form is
equal to
${\rm res}_w(\tr (\eta Dgg^{-1}))$.
The second component $T^*Jac(\Sigma)$ is the pair $(t,L)$, where $L$ is a point
of $Jac(\Sigma)$ and $t$ is the corresponding co-vector. The
canonical one-form is $\lan t,DL\ran_{Jac}$ and the brackets
denote the pairing between
vectors and co-vectors on the Jacobian.

The group  ${\cal G}_{out}$ acts as
$(g,\eta)\to (gh_{out},\eta)$. The corresponding
momentum constraint can be reformulated as
the following condition: $g\eta g^{-1}$ is
the restriction of some   Lie$^*(G)$-valued
 form on the complement to $w$.
The group  ${\cal G}_{int}$ acts as
$(g,\eta)\to (h_{int}g,h_{int}\eta h^{-1}_{int})$.
 The momentum constraint means
 that $\eta$ is holomorphic in ${\cal U}_w$ if $w$ is a
generic point, or it has the first order pole if $w$ is a marked
point.

\section{Symplectic Hecke correspondence}
\setcounter{equation}{0}
In this section we consider only $\GLN$-bundles.

\subsection{Hecke correspondence. }
Let $E$ and $\ti{E}$ be two bundles over $\Si$
of the same rank. Assume that there is a map
$\Xi^+~ \colon E\to \tilde{E}$ (more precisely
a map of the sheaves of sections $\G(E)\to\G\tilde{E}$)   such that
it is an isomorphism on the complement to
$w$ and it has one-dimensional cokernel at $w\in\Sigma$~:
\beq{aut}
0\to E\stackrel{\Xi^+}{\rightarrow}
\tilde{E}\to \Bbb C|_w\to 0\,.
\eq
It is the so-called {\sl upper modification} of the bundle $E$ at the point $w$..
On the complement to the point $w$ consider the map
$$
E\stackrel{~\Xi^-}\leftarrow\ti{E}\,,
$$
such that $\Xi^-\Xi^+=$Id. It defines {\sl the lower modification} $\gH_w^-$
at the point $w$.

\begin{defi}
The upper
 Hecke correspondence (HC) at the point $w\in\Si$  is
an \\
 auto-correspondence  ${\frak H}^+_w$
on the moduli space of Higgs bundles
${\cal H}$ related to the upper modification  $\Xi^+$ (\ref{aut}).
\end{defi}
HC ${\frak H}^+_w$ has components
placed only at ${\cal M}^{(d)}\times{\cal M}^{(d+1)}$.
The lower HC is defined in the similar way.
In this form the HC was used in the Hitchin systems in Ref.~\cite{K,ER}.

Now consider two quasi-parabolic bundles $E$ and $\ti{E}$ with the
flag structure at the marked points. While the flag $E_\al(s)$ at $w_\al$ corresponding
to $E$ has the form (\ref{fl}), for  $\ti{E}$ we declare the following flag
structure
$$
\ti{E}_\al(s)=\ti{Fl}_1(\al)\supset\cdots\supset
\ti{Fl}_{s_\al}(\al)\supset \ti{Fl}_{s_\al+1}(\al)=0\,,
$$
where $\ti{Fl}_{k}\sim Fl_{k-1}/Fl_{s_\al} $ for $s_\al+1\geq k \geq 2$.
We define $\ti{E}$  in terms of the sheaves of sections $\G(E)$.
Let $\Xi^+_\alpha$ be a map of the sheaves of sections $\G(E)\to\G(\tilde{E})$
such that
it is an isomorphism on the complement to a marked point $w_\alpha\in\Si$.
Let $\sigma\in\G(E)$ and
$\Xi^+_\alpha~:~\sigma\to\ti{\sigma}\in\G(\tilde{E})$.
If $\sigma|_{w_\al}\in Fl_{k-1}$,  then $\ti{\sigma}|_{w_\al}\in\ti{Fl_k}$.
The section $\sigma$ can be singular of order one if its principle part
belongs to $Fl_{s_\al}$.

All together means that $\Xi^+_\alpha$
 acts as the shift on the infinite flag (\ref{fl1}) at the marked point
\beq{aut1}
\Xi^+_\alpha ( LFl_{r,k})=LFl_{r-1,k}\,.
\eq
We call $\Xi^+_\alpha$ the upper modification of the quasi-parabolic bundle
$E$.
The lower modification of the quasi-parabolic bundles acts in the opposite
direction.
It looks like the upper modification (\ref{aut}), but we temporally do not
assume that $\Xi^+_\alpha$ has a one-dimensional cokernel.

\begin{defi}
The upper
 Hecke correspondence  of the quasi-parabolic bundles at the marked point $w_\al$  is
an auto-correspondence  ${\frak H}^+_w$
on ${\cal M}$ related to the the upper modification $\Xi^+$ (\ref{aut1}).
\end{defi}
Let the flag
$E_\al$ (\ref{fl}) has a one-dimensional subspace
($ \dim(Fl_{s_\al})=1$). In this case the
 upper modification $\Xi^+_\alpha$ can be fixed in the following way.
Let $(e_1,\ldots,e_N)$ be a basis
of local sections of $E$ compatible with the flag structure
$$
Fl_1\to(e_1,\ldots,e_N),\ldots,Fl_{s_\al}\to (e_N)\,.
$$
It follows from Definition 3.2 that
$\Xi_\al^+$
can be gauge transformed to the canonical form
\beq{hc+}
\Xi_\al^+=\mat{0}{{\rm Id}_{N-1}}{z_\al}{0}\,.
\eq
It is just the Coxeter transformation in the loop algebra $L(\gln)$,
that has been defined on the punctured disk $D^*_\al\subset{\cal U}_\al$
in Subsection 2.4.
 The Coxeter transformation provides the  upper modification
 $E^d\to\ti{E}^{d+1}$. In fact,
the sheaf of sections $\G(\tilde{E})$ coincides with the
sheaf of sections $\G(E)$ with a singularity of the first order at $w_\al$
and the singular section lies  in the kernel of $\Xi^+$ (see Ref.\cite{AL}).
For the local basis of $\G(E)$ we have
$(e_Nz^{-1},e_1 ,\ldots,e_{N-1})$. In this way the HC of the quasi-parabolic
bundles is described by the diagram (\ref{aut}).

In a similar way the lower modification can be transformed  to the form
\beq{hc-}
\Xi_\al^+=\mat{0}{z^{-1}_\al}{{\rm Id}_{N-1}}{0}\,.
\eq

\subsection{ Symplectic Hecke correspondence}.
We define a map of the Higgs bundles
$f~:~(E,\eta)\to(\ti{E},\ti{\eta})$ as the bundle map
$f~:~E\to\ti{E}$ such that
\beq{mh}
f\eta=\ti{\eta}f\,.
\eq
Consider two Higgs bundles $(E,\eta)$ and $(\ti{E},\ti{\eta})$,
where $E$ is a quasi-parabolic bundle
are in $\ti{E}$ is the upper modification $\Xi_\al^+$ of $E$  at  $w_a\in\Si$.
We call $(\ti{E},\ti{\eta})$ it the upper modification of $(E,\eta)$ if
$\Xi_\al^+\eta=\ti{\eta}\Xi_\al^+$.

\begin{defi}
 The upper symplectic Hecke correspondence (SHC) $\gS^+_\alpha$ at a point $w_\alpha$ is  an
auto-correspondence on $T^*\mathcal{M}$ related to the upper modification $\Xi_\al^+$
of the Higgs bundles.
\end{defi}
The lower SHC $\gS^-_a$ is defined in the similar way.

Let $w_\al$ be a marked point.
The  Higgs field $\eta$
 has the first order poles at $w_\al$ (\ref{pol}) and the residue $p_\al^{(0)}$
of $\eta$ defines an orbit
${\cal O}_\al$.

\begin{lem}
The  gauge transforms $\Xi^{\pm}_\al$ corresponding to $\gH^\pm_\al$ \\
$\bullet$ do not change singularity of the Higgs field at $w_\al$;\\
$\bullet$ are symplectic;\\
$\bullet$ preserve the  Hamiltonians (\ref{bb7}).
\end{lem}
{\sl Proof}. \\
The choice of $p_\al^{(0)}$ (\ref{orb}) is consistent with the canonical
forms (\ref{hc+}),(\ref{hc-}) of $\Xi^{\pm}_\al$ and their action
does not change the order of the pole.
The action is symplectic with respect to (\ref{sf})
since $\Xi^{\pm}_\al$ depends only on $p^{(0)}_\al$.
The invariance of the Hamiltonians follows from (\ref{mh}).
 $\Box$

In particular, Lemma 3.1 means that $\Xi^{\pm}_\al$ preserves the whole
Hitchin hierarchy defined by the set of Hamiltonians (\ref{bb7}) and the
symplectic form (\ref{sf}).

\subsection{SHC and skew-conormal bundles.}

Here we consider the curves without marked points.
The general case can be derived in the similar way and we drop it for the
sake of simplicity.

For any smooth correspondence $Z$ between equi-dimensional varieties
$X$ an $Y$ we
define a {\em skew-conormal
bundle} ${\cal SN}^*Z$ of $Z$ as follows.
Let
$$
\nu=(\nu_X,\nu_Y)
\in { T}^*_z(X\times Y)=
{ T}^*_xX\oplus { T}^*_y Y
$$
be a co-vector attached
to a point
$z=(x,y)\in Z\subset X\times Y$. It
belongs to the fiber ${\cal SN}_z^*Z$ of the skew-conormal bundle
${\cal SN}^*Z$ at the point $z$ iff
for any vector $v=(v_X,v_Y)$ tangent to $Z$
$$
\nu_X(v_X)=\nu_Y(v_Y)\,.
$$
Note that for the conormal
bundle one has the opposite sign:
$\nu_X(v_X)=-\nu_Y(v_Y)$.

The total space of the   skew-conormal
bundle is a Lagrangian subvariety of the
total space of the cotangent bundle
${ T}^*(X\times Y)$  with respect to the symplectic
form $\omega_X-\omega_Y$, where $\omega$ denotes
the canonical symplectic form on the cotangent
bundle.
 So, the skew-conormal bundle of a
correspondence is rather close
to the graph of a symplectic map between
cotangent bundles.

\begin{predl}
The graph of the SHC ${\frak S}_w$ is
isomorphic to the skew-conormal bundle
${\cal SN}^*{\frak H}_w$ of the
usual Hecke correspondence
${\frak H}_w$.
\end{predl}
{\sl Proof.}\\
As it was explained in Subsection 2.4, a $\GLN$-bundle $E$ is
determined
by the pair $(g,L)$ in a neighborhood of
a point $w\in \Si$.
An upper HC of $E$ corresponds to $(\ti{g},L)$, where $\tilde{g}=\Xi g$ and
\beq{hc}
\bp\Xi=0~{\rm in}~ {\cal U}_w,~~
{\rm ord}_w({\rm det}(\Xi))=1\,.
\eq
Therefore, the skew-conormal
 bundle ${\cal SN}^*{\frak H}_w$ of the
HC ${\cal SN}^*{\frak H}_w$ can be described by the  data
$$
(g,\tilde{g};\eta, \tilde{\eta};t,\ti{t},L),~
\tilde{g}=\Xi g\,,
$$
where $\Xi$ satisfies (\ref{hc}), and
\beq{hc0}
\lan t,DL\ran_{Jac}=\lan \ti{t},DL\ran_{Jac}
\eq
\beq{hc1}
{\rm res}_w({\rm Tr} (\tilde{\eta} D\tilde{g}\tilde{g}^{-1}))=
{\rm res}_w({\rm Tr} (\eta Dgg^{-1}))
\eq
for any variations of $g$ and $\tilde{g}$, that preserve properties of
$\Xi=g^{-1}\tilde{g}$.
The first condition (\ref{hc0}) means that $t=\tilde{t}$.

The condition (\ref{hc1}) can be rewritten as
$$
{\rm res}_w\left(
{\rm tr} (\tilde{\eta} D\Xi\Xi^{-1} +
(\Xi^{-1}\tilde{\eta} \Xi -\eta) g^{-1}D\,g)
\right)=0\,.
$$
Since variations of $g$ and $\Xi$ are independent,
 both terms in the last expression must vanish separately:
\beq{hc2}
 {\rm res}_w({\rm tr} (\tilde{\eta} D\Xi\Xi^{-1})) =0\,,
\eq
\beq{hc3}
{\rm res}_w\left({\rm tr}
(\Xi^{-1}\tilde{\eta} \Xi -\eta) g^{-1}D\,g)
\right)=0\,.
\eq

Consider first the case when $w$ is not a marked point.
Then we will demonstrate that (\ref{hc2}) means that
$\eta'=\Xi^{-1}\tilde{\eta}\Xi$
is holomorphic in $w$.
Consider the value of $\Xi$ at zero, this
matrix has rank $N-1$. Denote by $K$ its
kernel and by $I$ its image. An essential
variation of $\Xi$ corresponds to the variations
of its image, so it is a map $I\to {\Bbb C}^N/I$.
This variation corresponds to the right action:
$D\,\Xi=\Xi\epsilon $. The singular part
$\Xi^{-1}_{sing}$ of $\Xi^{-1}$ at $w$
is an operator of rank $1$.
Its kernel equals $I$ and its image equals
$K$, so the singular part $\eta_{sing}'$ of
$\eta'$ is a map
from
${\Bbb C}^N/{\rm Ker}(\Xi^{-1}_{sing})=
{\Bbb C}^N/I$ to ${\rm Im}(\Xi^{-1}_{sing})=I$.
The first condition can be rewritten as:
$$
0={\rm res}_w({\rm tr} (\tilde{\eta} D\Xi\Xi^{-1}))=
{\rm res}_w({\rm tr} (\eta' \Xi^{-1} D\Xi))=
{\rm tr}(\eta'_{sing}\epsilon)
$$
for any $\epsilon\in{\rm Hom}(I, {\Bbb C}^r/I)$.
The space
${\rm Hom}(I, {\Bbb C}^r/I)$ is dual to
${\rm Hom}( {\Bbb C}^r/I,I)$, so $\eta'_{sing}$
vanishes and $\eta'$ is holomorphic.

Note that $\eta'$ determines some Higgs field for $g$.
 Indeed, it is holomorphic in ${\cal U}_w$
and \mbox{$g^{-1}\eta'g=\tilde{g}^{-1}\tilde{\eta}
\tilde{g}$} is the restriction of some one-form
on the complement to $w$.
As the canonical one-form $\tr(\eta Dgg^{-1})$
is non-degenerate on $T^*{\cal M}_n^d$, from the second condition
(\ref{hc3}) we conclude that $\eta'=\eta$.

If $w$ is a marked point then $\Xi$ is fixed and it maps the Higgs field
$\eta$ into the Higgs field $\ti{\eta}$ (Lemma 3.1). There is no variation
of $\Xi$ and we immediately have that again $\eta'=\eta$. $\Box$

\subsection{B\"{a}cklund transformation.}
Consider the Higgs bundles with the quasi-parabolic structures at the marked
points.
The gauge transformations $\Xi^{\pm}_\al$ related to the SHC $\gS^\pm_\al$
depend only on the marked point $w_\al$.
They define the maps of the Hitchin systems
\beq{b.11}
\gS^+_\al\sim\xi^{\al}:
~T^*{\cal M}^{(d)}(\Si_n,G)\to T^*{\cal M}^{(d+1)}(\Si_n,G)\,,
\eq
\beq{b.12}
\gS^-_\al\sim\xi_{\be}:
 ~T^*{\cal M}^{(d)}(\Si_n,G)\to T^*{\cal M}^{(d-1)}(\Si_n,G)\,.
\eq
Consider  consecutive upper and lower modifications
\beq{b.13}
\xi^{\al_1}_{\al_2}=
\xi^{\al_1}\cdot\xi_{\al_2}\,.
\eq
Since ${\rm deg}(E)$ does not change it is a symplectic transform
$T^*{\cal M}_n^d(\Si,E)$. In this way $\xi^{\al_1}_{\al_2}$
maps solutions of the Hitchin hierarchy into solutions.
\begin{cor}
 The map (\ref{b.13}) is {\em the B\"{a}cklund transformation},
parameterized by  a pair of marked points $(w_{\al_1},w_{\al_2})$.
\end{cor}
We can generalize (\ref{b.13}) as
$$
\xi^{\al_{j_1};\ldots ;\al_{j_s} }
_{\al_{i_1};\ldots ;\al_{i_s}}=
\xi^{\al_{j_1}}\cdot\xi_{\al_{i_1}}\cdots\,.
$$
Because the B\"{a}cklund transformation is a canonical one we can consider
a discrete Hamiltonian system defined on the phase space
$T^*{\cal M}_n^d(\Si,E)$.
They pairwise commute and in terms of the angle variables generate
a lattice in the Liouville torus \cite{Ve,SK}. In our case
the dimension of the Liouville torus is equal to
$\dim{\cal M}_n^d$ (\ref{pb}), but the lattice we have constructed has
in general a smaller dimension.

Note that when $\Si_n$ is an elliptic curve,
the Hitchin systems corresponding
to $d=kN$ and $d=0$ ($d=$deg$(V)$) are
equivalent. Hence, in this case one can construct some
B\"{a}cklund transformations
by applying the upper SHC $N$ times.

\section {Elliptic CM system - elliptic $\SLN$-rotator correspondence}
\setcounter{equation}{0}

\subsection{ Elliptic CM system.}
The elliptic CM system was first introduced in the quantum version
\cite{C}. It is defined on the phase space
\beq{8.1}
{\mathcal R}^{CM}=\left(\bfv=(v_1,\ldots,v_N),~\bfu=(u_1,\ldots,u_N),
~~\sum_j v_j=0,~\sum_ju_j=0\right)\,,
\eq
with the canonical symplectic form
\beq{sf1}
\om^{CM}=(D\bfv\wedge D\bfu)\,.
\eq
The second order with respect to the momenta $\bfv$ Hamiltonian is
$$
H_2^{CM}=\oh\sum_{j=1}^Nv_j^2+\nu^2\sum_{j>k}\wp(u_j-u_k;\tau)\,.
$$

It was established in \cite{Ma,GN} that the elliptic CM system
can be derived in the Hitchin approach.
The Lax operator  $L^{CM}$ is the reduced Higgs field $\eta$ over
the elliptic curve
$$
E_\tau=\mC/\mL\,,~~\mL=\mZ+\tau\mZ
$$
 with a marked point $z=0$.
In this way the phase space ${\mathcal R}^{CM}$ is the space of pairs
\begin{center}

(quasi-parabolic $SL_N$-bundle $V$ over $E_\tau$, the Higgs field
$L^{CM}$ on this bundle (\ref{8.2})).

\end{center}
The bundle is determined  by the transition
functions (the multipliers)
\beq{b.14}
Id_N:~~z\to z+1\,,
\eq
$$
\bfe(\bfu)=\di(\bfe(u_1),\ldots,\bfe(u_N)):~~z\to z+\tau\,,
$$
where $\bfe$ is defined in (\ref{AA0}).
The Lax operator $L^{CM}(z)$ is the quasi-periodic one-form
\beq{b.15}
L^{CM}(z+1)=L^{CM}(z),~~
L^{CM}(z+\tau)=\bfe(-\bfu)L^{CM}(z)\bfe(\bfu)\,.
\eq
It is the $N$-th order matrix with the first order pole at
$z=0$ and the residue
\beq{8.2b}
p^{(0)}={\rm Res}_{z=0}(L^{CM}(z))=L_{-1}^{CM}=\nu
\left(
\begin{array}{cccc}
0&1&\cdots&1\\
1&0&\cdots&1\\
\vdots&\vdots&\ddots&\vdots\\
1&1&\cdots&0
\end{array}
\right),
\eq
This residue defines the minimal coadjoint orbit $\mathcal{O}$
(\ref{tor}) $(\dim({\cal O})=2N-2)$. These  degrees of freedom
are gauged away by the action of rest gauge symmetries generated by the
constant diagonal matrices. For this reason the second term in (\ref{sf})
does not contribute in the symplectic form (\ref{sf1}).

The column-vector $e_1=(1,1,\cdots, 1)$ is an eigen-vector $e_1$
\beq{e1}
L_{-1}^{CM}e_1=(N-1)\nu e_1.
\eq
There is also an $(N-1)$-dimensional eigen-subspace  $T_{N-1}$ corresponding
to the degenerate eigen-value $-\nu$
\beq{8.2a}
L_{-1}^{CM}e_\bfa=-\nu e_\bfa,~~e_\bfa=(a_1,\ldots,a_N),~(\sum_na_n=0)\,.
\eq

The quasi-periodicity (\ref{8.2b}) leads to the following form
of $L^{CM}$
\beq{8.2}
L^{CM}=P+X,~{\rm where}~P=\di(v_1,\ldots,v_N),~~X_{jk}=\nu\phi(u_j-u_k,z)\,,
\eq
and $\phi$ is defined as (\ref{A.3}).

 The $M^{CM}$-operator corresponding to $H^{CM}_2$
 has the form
\beq{8.3}
M^{CM}=-D+Y,~{\rm where}~D=\di(Z_1,\ldots,Z_N),~~Y_{jk}=y(u_j-u_k,z)\,,
\eq
$$
Z_j=\sum_{k\neq j}\wp(u_j-u_k) ,~~y(u,z)=\frac{\p \phi(u,z)}{\p u},.
$$

\subsection{The elliptic $\SLN$-rotator.}
The elliptic  $\SLN$-rotator  is an example of the  Euler-Arnold top \cite{Ar}.
It is defined on
 a coadjoint orbit of $\SLN$:
\beq{8.4}
{\mathcal R}^{rot}=\{\bfS\in\sln,~~\bfS=g^{-1}\bfS^{(0)}g\}\,,
\eq
where $g$ is defined up to the left multiplication on
the stationary subgroup $G_0$ of $\bfS^{(0)}$.
The phase space ${\mathcal R}^{rot}$ is equipped with the
Kirillov-Kostant symplectic form
\beq{sf2}
\om^{rot}=\tr(\bfS^{(0)}Dgg^{-1}Dgg^{-1})\,.
\eq
 The Hamiltonian is defined as
\beq{8.5}
H^{rot}=-\frac{1}{2}\tr(\bfS J(\bfS))\,,
\eq
where $J$ is a linear operator on  Lie$(\SLN)$. The inverse
operator is called the inertia tensor.
The equation of motion takes the form
\beq{em}
\p_t\bfS=[J(\bfS),\bfS].
\eq

 We consider here a special form $J$, that provides the integrability
of the system.
Let
$$
J(\bfS)=\bfJ\cdot\bfS=\sum_{mn}J_{mn}S_{mn}\,,
$$
 where $\bfJ$ is a $N$-th order matrix
\beq{8.6}
\bfJ=\{J_{mn}\}=
\left\{\wp\left[\begin{array}{c}m\\n\end{array}\right]\right\}\,,~~
(m,n=1,\ldots,N)\,,~~(m,n\in\mZ\,~{\rm mod}~ N,~m+n\tau\not\in\mL)\,,
\eq
$$
\wp\left[\begin{array}{c}m\\n\end{array}\right]=
\wp(\frac{m+n\tau}{N};\tau)\,.
$$
We write down (\ref{em}) in the basis of the sin-algebra $\bfS=S_{mn}E_{mn}$
(see (\ref{AA3}))
\beq{8.5a}
\p_tS_{mn}=\frac{N}{\pi}\sum_{k,l}S_{k,l}S_{m-k,n-l}
\wp\left[\begin{array}{c}k\\l\end{array}\right]
\sin\frac{\pi}{N}(kn-ml)\,.
\eq

The elliptic rotator is a Hitchin system \cite{Ma}.
We give a proof of this statement.
\begin{lem}
The elliptic
$\SLN$-rotator is a Hitchin system corresponding to the Higgs
quasi-parabolic $\GLN$-bundle $E$ $\ \ \ \ $
(deg$(E)$=$1$)
over the elliptic curve $E_\tau$ with the marked point $z=0$.
\end{lem}
{\sl Proof.}

It can be proved that (\ref{8.5a}) is equivalent to the Lax equation.
The Lax matrices in the basis of the sin-algebra take the form
\beq{8.7}
L^{rot}=\sum\limits_{m,n}
S_{mn}\varphi\left[\begin{array}{c}m\\n\end{array}\right](z)
E_{mn},~~~
\varphi\left[\begin{array}{c}m\\n\end{array}\right](z)=
\bfe(-\frac{nz}{N})\phi(-\frac{m+n\tau}{N};z)\,,
\eq
\beq{8.8}
M^{rot}=\sum\limits_{m,n}
S_{mn}f\left[\begin{array}{c}m\\n\end{array}\right](z)E_{mn},~~
f\left[\begin{array}{c}m\\n\end{array}\right](z)=
\bfe(-\frac{nz}{N})\partial_{u}\phi(u;z)|_{u=-\frac{m+n\tau}{N}}\,.
\eq
They lead to the Lax equation for the matrix elements
$$
\p_tS_{mn}\varphi\left[\begin{array}{c}m\\n\end{array}\right](z)=\sqrt{-1}
\sum_{k,l}
S_{m-k,n-l}S_{kl}\varphi\left[\begin{array}{c}m-k\\n-l\end{array}\right](z)
f\left[\begin{array}{c}k\\l\end{array}\right](z)
\sin\frac{\pi}{N}(nk-ml)\,.
$$
Using the Calogero functional equation (\ref{ad2})
we rewrite it in the form (\ref{8.5a}).
Since
$$
\frac{1}{N}\tr(L^{rot})^2=-2H^{rot}+\tr\bfS^2\wp(z)\,,
$$
$H^{rot}$ is the Hitchin quadratic integral.

The Lax operator satisfies the Hitchin equation
$$
\bp L^{rot}=0,~~ {\rm Res}L^{rot}|_{z=0}=2\pi\sqrt{-1}\bfS
$$
and is quasi-periodic
\beq{8.9}
L^{rot}(z+1)=Q(\tau)L^{rot}(z)Q^{-1}(\tau),
\eq
\beq{8.10}
L^{rot}(z+\tau)=\ti{\La}(z,\tau)L^{rot}(z)(\ti{\La}(z,\tau))^{-1},
\eq
where $\ti{\La}(z,\tau)=-\bfe(\frac{-z-\oh\tau}{N})\La$ and the matrices
$Q$ and $\La$ are defined in (\ref{AA11}),(\ref{AA2}).
The transition functions
\beqn{8.11}
Q(\tau):~~z\to z+1, \\
\ti{\La}(z,\tau):~~z\to z+\tau
\eqn
 define $\GLN$-bundle over $E_\tau$ with deg$(V)=1$.
For these bundles we have $\dim({\cal M}_0^1)=1$ (\ref{2.0}) and
 after the symplectic
reduction we come to the coadjoint orbit $G_0\setminus \SLN$  (\ref{8.4}).
 The
Kirillov-Kostant
form (\ref{sf2}) arises as the last terms in (\ref{sf})
attributed to the point $z=0$.
Thus, the phase space of the SL$_N$-rotator is
the space of the Higgs fields $L^{rot}$ on the bundle determined
by multipliers $Q$, $\tilde\Lambda$  with the
first order singularities at zero.
$\Box$

\subsection{A map ${\mathcal R}^{CM}\to {\mathcal R}^{rot}$.}

We  construct  a map from the phase space of the elliptic
 CM system  ${\mathcal R}^{CM}$
 into the phase space of the $SL_N$-rotator
${\mathcal R}^{rot}$. We assume here that the $SL_N$-rotator is living on the most degenerate
orbit corresponding to $L_{-1}^{CM}$ (\ref{8.2b}).
The phase space of CM systems with spins is mapped into the general
coadjoint orbits. This generalization is straightforward.
In this way, for $N=2$ we describe the upper horizontal arrow in Fig.(1.1).

The  map is defined as the conjugation of $L^{CM}$ by
some matrix $\Xi(z)$:
\beq{8.11a}
L^{rot}=\Xi\times L^{CM}\times \Xi^{-1}.
\eq
It follows from comparing (\ref{b.15}) with (\ref{8.9}) and (\ref{8.10}) that
$\Xi$ must intertwine the multipliers of bundles:
  \beq{8.12}
\Xi(z+1,\tau)= Q\times \Xi(z,\tau)\,,
\eq
\beq{8.13}
\Xi(z+\tau,\tau)=\tilde\Lambda(z,\tau)\times
\Xi(z,\tau)
\times{\rm diag}
({\bf e}(u_j))\,.
\eq
The matrix $\Xi(z)$ degenerates at $z=0$, and the column-vector
$(1,\cdots,1)$, in accordance with Lemma 3.1, should belong to the kernel
 of $\Xi(0)$.
In this case,
$\Xi\times L^{CM}\times \Xi^{-1}$ has a  first order pole at $z=0$.

Consider the following $(N\times N)$-
matrix $\tilde\Xi(z, u_1,\ldots,u_N;\tau)$~:
\beq{8.14}
\tilde\Xi_{ij}(z, u_1,\ldots,u_N;\tau) =
\theta{\left[\begin{array}{c}
\frac{i}N-\frac12\\
\frac{N}2
\end{array}
\right]}(z-Nu_j, N\tau ),
\end{equation}
where $\theta{\left[\begin{array}{c}
a\\
b\end{array}
\right]}(z, \tau )$ is the theta function with a characteristic (\ref{A.30}).
Sometimes we omit nonessential arguments of $\Xi$ for brevity.

\begin{lem}
 The matrix $\tilde\Xi$ is transformed under the translations $z\to z+1$,
$z\to z+\tau$ and $u_j\to u_j+1$,
$u_j\to u_j+\tau$  as~:
\beq{8.15}
\tilde\Xi(z+1,\tau)= -Q\times \tilde\Xi(z,\tau)\,,
\eq
\beq{8.16}
\tilde\Xi(z+\tau,\tau)= \tilde{\Lambda}(z,\tau)\times \tilde
\Xi(z,\tau)
\times{\rm diag}
({\bf e}(u_j)),
\eq
$$
\tilde{\Lambda}(z,\tau)=-{\bf e}\left(-\frac{\tau}{2N}-\frac{z}N\right)\La\,;
$$
\beq{8.17}
\tilde\Xi( u_j+1,;\tau)=
\tilde\Xi( u_j;\tau)\times {\rm diag}(1,\cdots, (-1)^N,\cdots, 1)\,,
\eq
\beq{8.18}
\tilde\Xi( u_j+\tau;\tau)=
\tilde\Xi( u_j;\tau)\times{\rm diag}(1,\cdots, (-1)^N
{\bf e}\left(-\frac{N\tau}2+z-Nu_j\right)\,\cdots, 1)\,.
\eq
\end{lem}
{\sl Proof}.\\
The statement of the lemma follows from the properties of the
theta functions with
characteristics (\ref{A.31})-(\ref{A.33}). $\Box$

Now we assume that $\sum u_j=0$, so $u_N$ is no more an independent
variable, but it is equal to $-\sum_{j=1}^{N-1}u_j$.

The determinant formula of the Vandermonde type \cite{Has}
\beq{determ}
\det\left[\frac{\tilde\Xi_{ij}(z, u_1,\ldots,u_N;\tau)}
{\sqrt{-1}\eta(\tau)}\right]=
\frac{\vth(z)}{\sqrt{-1}\eta(\tau)}\prod\limits_{1\leq k<l\leq N}
\frac{\vth(u_l-u_k)}{\sqrt{-1}\eta(\tau)}
\eq
is used to show that the matrix $\tilde\Xi_{ij}(z)$ truly
degenerates at $z=0$. Here $\eta(\tau)$ is the Dedekind function.

\bigskip
\begin{lem}
The kernel of $\tilde\Xi$ at $z=0$ is generated
by the following column-vector~:
$$
\left\{(-1)^l \prod_{j<k;j,k\ne l}
\vartheta(u_k-u_j,\tau)\right\}, \quad l=1,2,\cdots,N.
$$
\end{lem}
{\sl  Proof}.\\
 We must prove that for any $i$ the following expression
\beq{8.21}
\sum_{l=1}^N
(-1)^l
\theta{\left[\begin{array}{c}
\frac{i}N-\frac12\\
\frac{N}2
\end{array}
\right]}(z-Nu_l, N\tau )
\prod_{j<k;j,k\ne l}
\vartheta(u_k-u_j,\tau)
\eq
vanishes.
 First, the symmetric group $S_N$ acts on $\bfu$ by permutation
of $u_1,\ldots ,u_N$ and (\ref{8.21}) is antisymmetric with respect to the
 $S_N$ action.
 Hence it vanishes on the hyperplanes $u_i=u_j$.
As a function on $u_1$,  (\ref{8.21})   has $2N$ zeroes:
$N-2$ zeroes $u_1=u_k$, $k\ne 1,N$, $N-2$ zeroes $u_N=u_k$,
 $k\ne 1,N$
and four zeroes $u_1=u_N$ (the last equation is $2u_1=
-\sum_{j=2}^{N-1}u_j$).

Second, (\ref{8.21})  is quasiperiodic
 with respect to the shifts $u_1\to u_1+1$, $u_1\to u_1+\tau$
with multiplicators $1$ and ${\bf e}\left(-(N-1)\tau -(N-1)(u_1-u_N)\right)$.
Any quasiperiodic function with such multiplicators is either zero
or has $2N-2$ zeroes. Since our expression vanishes in $2N$ points
it vanishes identically.  $\Box$

It follows from the previous lemmas that the matrix
\beq{xi1}
\Xi(z)=\tilde\Xi(z)\times{\rm diag}\left((-1)^l
\prod_{j<k;j,k\ne l}
\vartheta(u_k-u_j,\tau)\right)
\eq
is the singular gauge transform from Lemma 2.1 that
maps $L^{CM}$ to $L^{rot}$.
This transformation leads to the symplectic map
\beq{map}
{\mathcal R}^{CM}\to {\mathcal R}^{rot},~~
(\bfv,\bfu)\mapsto \bfS.
\eq
Consider in detail the case $N=2$.
Let
$$
\bfS=S_a\si_a,
$$
where $\si_a$ denote the sigma matrices subject to the commutation
relations
$$
[\si_a,\si_b]=2\sqrt{-1}\varepsilon_{abc}\si_c\,.
$$
Then the transformation has the form
\beq{S}
\left\{
\begin{array}{l}
S_1=-v\frac{\theta_{10}(0)}{\vartheta'(0)}
\frac{\theta_{10}(2u)}{\vartheta(2u)}-
\nu\frac{\theta_{10}^2(0)}{\theta_{00}(0)\theta_{01}(0)}
\frac{\theta_{00}(2u)\theta_{01}(2u)}{\vartheta^2(2u)}\,,
 \\
S_2=-v\frac{\theta_{00}(0)}{\sqrt{-1}\vartheta'(0)}
\frac{\theta_{00}(2u)}{\vartheta(2u)}-
\nu\frac{\theta_{00}^2(0)}{\sqrt{-1}\theta_{10}(0)\theta_{01}(0)}
\frac{\theta_{10}(2u)\theta_{01}(2u)}{\vartheta^2(2u)}\,,
\\
S_3=-v\frac{\theta_{01}(0)}{\vartheta'(0)}
\frac{\theta_{01}(2u)}{\vartheta(2u)}-
\nu\frac{\theta_{01}^2(0)}{\theta_{00}(0)\theta_{10}(0)}
\frac{\theta_{00}(2u)\theta_{10}(2u)}{\vartheta^2(2u)}\,.
 \\
\end{array}
\right.
\eq
Formulae of this kind were obtained in \cite{FT}.

\subsection{ B\"{a}cklund transformations in the CM systems.}

We now use the map (\ref{map}) to construct the B\"{a}cklund
transformation in the CM systems
 $$
\xi:~(\bfv,\bfu)\to(\ti{\bfv},\ti{\bfu})\,.
$$
Let the Lax matrix depends on the new coordinates and momenta
$L=L(\ti{\bfv},\ti{\bfu})$.
Consider the upper modification $\Xi(z)$ (\ref{xi1}).
To construct the B\"{a}cklund transformation $\xi$,  we map
$(\bfv,\bfu)$ and
$(\ti{\bfv},\ti{\bfu})$ to the same point $\bfS\in {\mathcal R}^{rot}$:
\bigskip
$$
\def\mapright#1{\smash{
    \mathop{\longrightarrow}\limits^{#1}}}
\begin{array}{rcccl}
           &                   &
\fbox{$L^{rot}(\bfS)$}&                           &     \\
           &\nearrow{\Xi(z,\bfu)}
&      &\nwarrow{\Xi(z,\ti{\bfu})}   &     \\
 \fbox{$L^{CM}(\bfv,\bfu)$}&  &\mapright{\xi}
& &\fbox{$L^{CM}(\ti{\bfv},\ti{\bfu})$}\\
\end{array}
$$
In this way we reproduce implicitly the general formula (\ref{b.13}) for
the B\"{a}cklund transformations. This transformation defines an integrable
discrete time dynamics of a CM system. One example of this dicretization
was proposed in \cite{NP}. It can be supposed to correspond to
$\xi$.

Another way to construct new solutions from $(\bfv,\bfu)$ is to act
by $N$ consecutive upper modifications
\beq{sbt}
\Xi(N)=D_N\Xi_N\cdots\Xi_j\cdots\Xi_2\cdot\Xi \,.
\eq
Here  the matrices $\Xi_j,~j=2,\ldots,N,$ satisfy the quasi-periodicity
conditions
$$
\Xi_j(z+\tau)=\ti{\La}^j\Xi_j(z) \ti{\La}^{1-j}\,,
$$
and $D_N$ is an arbitrary diagonal matrix.
We come back to the $N$-dimensional moduli space ${\cal M}^{(N)}$
(see (2.5)) and  to the map
$$
\def\mapright#1{\smash{
    \mathop{\longrightarrow}\limits^{#1}}}
L^{CM}(\bfv,\bfu)~\mapright{\Xi(N)}~L^{CM}(\ti{\bfv},\ti{\bfu})\,.
$$
If we break the chain (\ref{sbt}) on a step $k<L$, then we obtain the map
$$
L^{CM}\to L^{rot,k}\,,
$$
where $L^{rot,k}$ is the Lax operator for the elliptic rotator related to
the holomorphic bundle of degree $k$. It satisfies the quasi-periodicity
condition (\ref{8.9}) and
$$
L^{rot,k}(z+\tau) =\ti{\La}^jL^{rot,k}(z)\ti{\La}^{-j}
$$
instead of (\ref{8.10}).

\section{Hitchin systems of infinite rank}
\setcounter{equation}{0}
Here we generalize the derivation of finite-dimensional integrable
systems in the form (\ref{ab1})-(\ref{ab4}) on two-dimensional
integrable field theories.

\subsection{ Holomorphic $\hat{L}(\GLN)$-bundles.}

Let $L(\gln)$ be the loop algebra of ${C^\infty}$-maps
 $L(\gln):~S^1\to\gln$,
and $\hat{L}(\gln)$ be its central extension with the multiplication
\beq{ml}
(g,c)\times(g',c')=\left(gg',cc'\exp {\mathcal C}(g,g')\right),
\eq
where $\exp {\mathcal C}(g,g')$ is a 2-cocycle of $\hat{L}(\GLN)$
 providing the associativity of the multiplication.

Consider a holomorphic vector bundle $V$ of an infinite rank
over a Riemann curve $\Sigma_n$ with $n$ marked points. The bundle is
defined by the transition functions from $\hat{L}(\GLN)$.
Its fibers are isomorphic to the Lie algebra  $\hat{L}(\gln)$.
The holomorphic structure on $V$ is defined by the operator
$$
d^{''}:
\Omega_{C^\infty}^{(0)}(\Sigma_n,{\rm End}~V)\to
\Omega_{C^\infty}^{(0,1)}(\Sigma_n,{\rm End}~V)\,.
$$
It has two components $d^{''}=d_{\bA}^{''}+d_{\la}^{''}$. The first
component is
$$
d_{\bA}^{''}:
\Omega_{C^\infty}^{(0)}(\Sigma_n,L(\gln))\to
\Omega_{C^\infty}^{(0,1)}(\Sigma_n,L(\gln))\,.
$$
Locally
 $$
d_{\bA}^{''}=\bp+\bar{A},~~\bp=\p_{\bz},~~\bar{A}=\bA(x,z,\bz), ~x\in S^1\,.
$$
The operator $d_{\bA}^{''}$ acts on a $N$-dimensional column vector
$\vec{e}(x;z,\bz)$.
The second component is defined by the connection $d_{\la}^{''}$ on a trivial
linear bundle ${\cal L}$ on $\Sigma_n$, given by
$$
d_{\la}^{''}=\bp+\la\,.
$$
The field $\la$ is a map from $\Si_n$ to the central element of the Lie
algebra $\hat{L}(\gln)$.
A local section $\sigma$ of $V$ is holomorphic if
$d^{''}\sigma=0$. The sections allow to define
the transition functions.
We assume that $\bA$ and $\lambda$ are smooth at the marked points.

In addition we define  $n$ copies of the central extended loop groups
located at  the marked points
$$
\hat{L}G_\al=(g_\al(x),c_\al),~G_\al=\GLN,
 ~(\al=1,\ldots,n),~x\in S^1,
$$
with the multiplication (\ref{ml}).

Thus, we have the set $\mathcal{R}$ of fields playing the role of
the "coordinate space":
\beq{1.1}
{\mathcal R}=\left\{\bA,\lambda,(g_1,c_1),\ldots,(g_n,c_n)\right\}.
\eq

\subsection{ Gauge symmetries.}
Let ${\mathcal G}$ be the group of automorphisms of $\mathcal{R}$
(the gauge group).
$$
{\cal G}=C^\infty{\rm Map}(\Si_n\to \hat{L}(\GLN))=
\{f(z,\bz,x),s(z,\bz)\}\,,
$$
where $f(z,\bz,x)$ takes values in $\GLN$,
and $s(z,\bz)$ is the map to the central element of $\hat{L}(\GLN))$.
The
multiplication is pointwise with respect to $\Sigma_n$
$$
(f_1,s_1)\times(f_2,s_2)=\left(f_1f_2,s_1s_2\exp {\cal C}(f_1,f_2)\right)\,,
$$
where $\exp {\cal C}(f_1,f_2)$ is a map from $\Si_n$ to the 2-cocycle of
$\hat{L}(\GLN)$.

Let $(f_\al=f_\al(x),s_\al)$ be the value of the gauge fields at the marked
point $w_\al$.
The action of ${\cal G}$ on $\mathcal{R}$ takes the following form
\beq{1.2o}
\bA\to f^{-1}\bp f+f^{-1}\bA f,
\eq
\beq{1.2}
\la\to \la+s^{-1}\bp s+\oint\tr(\bA f^{-1}\p_xf)dx,
\eq
\beq{1.2a}
c_\al\to c_\al s_\al,~g_\al\to g_\al f_\al.
\eq
The quotient space ${\cal N}=\mathcal{R}/{\cal G}$
is the moduli space of infinite rank holomorphic bundles over Riemann
curves with marked points.

\subsection{ Phase space.}

The cotangent space to $\mathcal{R}$ has the
following structure. Consider  the analog of the Higgs field
$~\Phi\in\Omega_{C^\infty}^{(1,0)}(\Sigma_n,({\rm End}~V)^*)$. It is
a one-form $\Phi$ on $\Sigma_n$ taking values in the Lie coalgebra
$L^*(\gln)$.   Let $k$ be a scalar one-form
on $\Sigma_n$, $k\in\Omega_{C^\infty}^{(1,0)}(\Sigma_n)$. It is dual to the
field $\lambda$.
At the marked points we have the Lie coalgebras Lie$^*(G_\al)\sim L(\gln)$
along with the central elements $r_\al$, dual to $c_\al$.
Thus the cotangent bundle $T^*\mathcal{R}$ contains the fields
\beq{1.3}
T^*{\mathcal R}=\left\{(\bar{A},\Phi),(\lambda,k);
(g_1,c_1;p_1,r_1),\ldots,(g_n,c_n;p_n,r_n)\right\}\,.
\eq

There is a canonical symplectic structure on $T^*\mathcal{R}$.
For $F\in\Omega_{C^\infty}^{(1,0)}(\Sigma_n,({\rm End} ~V)^*)$ and
$G\in\Omega_{C^\infty}^{(0,1)}(\Sigma_n,\hat{L}(\gln))$ defines the pairing
$$
<F|G>=\int_{\Sigma_n}\oint \tr(FG){dx}.
$$
Then
\beq{1.4}
\omega=<D\Phi\wedge D\bA>+
\int_{\Sigma_n}D k\wedge  D\lambda+\sum_{\al=1}^n\omega_\al\,,
\eq
where $\omega_\al$ is a canonical form on $T^*\hat{L}(G_\al)$.
It is constructed in the canonical way by means of the Maurer-Cartan
form on $\hat{L}(G_\al)=\{g_\al,c_\al\}$. The result is
\beq{1.5}
\omega_\al=
\oint _{S^1_\al}\tr(D(p_\al g_\al^{-1})Dg_\al)+
D(r_\al c_\al^{-1})D c_\al
+\frac{r_\al}{2}\oint_{S^1_\al}
\tr\left(g_\al^{-1}D g_\al \p_x(g_\al^{-1}Dg_\al )\right).
\eq
\subsection{ Symplectic reduction.}

Now consider the lift of $\mathcal{G}$ to the global canonical
transformations
of $T^*{\mathcal R}$. In addition to (\ref{1.2o}),(\ref{1.2}),(\ref{1.2a})
we have the following action of $\mathcal{G}$
\beq{1.8}
\Phi\to f^{-1}k\p_x f+f^{-1}\Phi f,  ~~k\to k,
\eq
\beq{1.8a}
p_\al\to f_\al^{-1}p_\al f_\al+r_\al f_\al^{-1}\p_xf_\al,~~
r_\al\to r_\al\,.
\eq
This transformation leads to the moment map from the phase space
to the Lie coalgebra of the gauge group
$\mu: ~T^*{\mathcal R}\to {\rm Lie}^*({\mathcal G})$.
It takes the form
\beq{1.9}
\mu=\left(
\bp\Phi-k\p_x\bA+[\bA,\Phi]+\sum_{\al=1}^n
p_\al\de(z_\al),~
\bp k+\sum_{\al=1}^nr_\al\de(z_\al)
\right).
\eq
We assume that $\mu=(0,0)$. Therefore, we have the two holomorphity
conditions
\beq{1.10}
\bp\Phi-k\p_x\bA+[\bA,\Phi]+\sum_{\al=1}^n
p_\al\de(z_\al)=0\,,
\eq
\beq{1.11}
\bp k+\sum_{\al=1}^nr_\al\de(z_\al)=0\,.
\eq
 The constraint equation (\ref{1.11}) means that the $k$-component
 of the Higgs field
is a holomorphic one-form on $\Si$ with first order poles
at the marked points.

Let us fix a gauge
\beq{1.12}
\bL=f^{-1}\bp f+f^{-1}\bA f.
\eq
The same gauge action transform $\Phi$ as
\beq{1.13}
L=kf^{-1}\p_xf+f^{-1}\Phi f.
\eq
We preserve the same notations $g_\al,p_\al$ for the gauge transformed
variables.
The moment constraint equation (\ref{1.10}) has the same form in terms
of $\bL$ and $L$
\beq{1.14}
\bp L-k\p_x\bL+[\bL,L]+\sum_{\al=1}^n
p_\al\de(z_\al)=0\,.
\eq
Solutions of this equation along with (\ref{1.11}) define the
reduced phase space
$$
T^*{\cal R}//{\mathcal G}\sim T^*{\mathcal N}\,.
$$
The symplectic form (\ref{1.4}) on $T^*{\mathcal N}$ becomes
\beq{1.4a}
\omega=<\delta L|\delta\bL>+
\int_{\Sigma_n}\delta k\delta\lambda+\sum_{\al=1}^n\omega_\al\,.
\eq

\subsection{ Coadjoint orbits.}

 Consider in detail the symplectic form $\om$
(\ref{1.5}) on
$T^*\hat{L}(G)\sim \{(p,r);(g,c)\}$.
We omit the subscript $_\al$ below.
The following canonical transformation of $\om$  by
$(f,s)\in\hat{L}(G)$, where $s$ is a central element,
\beq{1.6}
g\to fg,~p\to p,~r\to r,~c\to sc, ~~f\in L(G)\,,
\eq
 has not been used so far.
The symplectic reduction with respect to this transformation leads to
the coadjoint orbits of $\hat{L}(\GLN)$.
In fact, the moment map
$$
\mu:T^*\hat{L}(G)\rightarrow{\rm Lie}^*(\hat{L}(\GLN))
$$
takes the form
$$
\mu=(-gpg^{-1}+r\p_xgg^{-1},r).
$$
Let us fix the moment $\mu=(p^{(0)},r^{(0)})$.
The result of the symplectic reduction of $T^*\hat{L}(G)$ is the
coadjoint orbit
$$
 {\mathcal O}(p^{(0)},r^{(0)})=
( p=-g^{-1}p^{(0)}g-r^{(0)}g^{-1}\p_xg,~r^{(0)})
=\mu^{-1}\left(T^*\hat{L}(\SLN)\right)/G_0,
$$
where  $G_0$ is the subgroup of $\hat{L}(\GLN)$ that preserves $\mu$
$$
G^0=\{g\in L(\GLN, ~s~{\rm is~arbitrary})~|~p^{(0)}=
-g^{-1}p^{(0)}g+r^{(0)}g^{-1}\p_xg\}.
$$

The symplectic form (\ref{1.5}) being pushed forward on $ {\mathcal O}$
takes the form
\beq{1.6a}
\om=\oint \tr(D (p g^{-1})D g)
+\frac{r^{(0)}}{2}\oint \tr\left(g^{-1}D gD  (g^{-1}\p_xg )\right).
\eq

In what follows we will consider the collection of
the orbits ${\mathcal O}_\al(p^{(0)}_\al,r^{(0}_\al)$ at the marked points
instead of the cotangent bundles $T^*\hat{L}(G_\al)$.
In this way we come to the notion of the Higgs bundle of infinite rank
$\hat{\cal H}_n^d$ (see (\ref{hb}) and (\ref{1.3}))
\beq{1.6b}
\hat{\cal H}_n^d=\left\{(\bar{A},\Phi),(\lambda,k),
{\mathcal O}_1(p^{(0)}_1,r^{(0}_1),
\ldots,{\mathcal O}_n(p^{(0)}_n,r^{(0}_n)\right\}\,.
\eq

\subsection{ Conservation laws I}.

The Higgs field $\Phi$ is transformed as a connection with respect to
the circles $S^1$ (\ref{1.8}). If the central charge $k\neq 0$, the standard
Hitchin integrals (\ref{ab4}) cease to be gauge invariant.
Invariant integrals are generated by the traces of the monodromies of the
Higgs field $\Phi$. The generating function of Hamiltonians is given by
\beq{1.15}
H(z)=\tr \left(P\exp\f1{k}\oint_{S_1}\Phi\right)\,,
\eq
where $z$ is a local coordinate of an arbitrary point.
At a marked point, $\Phi$ has a first order pole and
\beq{1.16}
H(z)=\sum_{j=-1}^{+\infty}H_jz^j\,.
\eq
Since $H(z)$ is gauge invariant one can replace $\Phi$ by $L$ in (\ref{1.15})
\beq{1.17}
H(z)=\tr \left(P\exp\f1{k}\oint_{S_1}L\right)\,.
\eq

\subsection{ Equations of motion.}

Consider the equations of motion on the "upstairs" space $T^*{\mathcal R}$
(\ref{1.6b}).
They are derived by means of the symplectic form $\om$ (\ref{1.4}),
where $\om_\al$ is replaced by (\ref{1.6a}), and the Hamiltonians
(\ref{1.15}), (\ref{1.16}).
Let $t_j$ be a time variable corresponding to the Hamiltonian $H_j$.
Taking into account
that $H_j$ is a functional depending on the Higgs field  and the central
charge $k$ only,  we arrive to the following  free system
\beq{1.20}
\p_j\Phi=0,
\eq
\beq{1.21}
\p_j\bA=\frac{\de H_j}{\de \Phi}\,,
\eq
\beq{1.22}
\p_j k=0,~~\p_j\la=\frac{\de H_j}{\de k},~~\p_jp_\al=0\,.
\eq

After the symplectic reduction we are led to the fields $\bL$ (\ref{1.12})
and $L$ (\ref{1.13}). For simplicity,
we keep the same notation for the coadjoint orbits variables $p_\al$,
so they are transformed as in (\ref{1.8a}).
Substituting (\ref{1.13}) in (\ref{1.20}) we obtain the Zakharov-Shabat
equation
\beq{1.23}
\p_jL-k\p_xM_j+[M_j,L]=0\,,~~(\p_j=\p_{t_j})\,,
\eq
where $M_j=\p_jff^{-1}$. The operator $M_j$ can be restored partly from the
second equation (\ref{1.21})
\beq{1.24}
\bp M_j-\p_j\bL+[M_j,\bL]=\frac{\de H_j}{\de L}\,.
\eq
The last two equations along with the moment constraint equation (\ref{1.14})
are the consistency conditions for the linear system
\beq{1.25}
(k\p_x+L)\Psi=0\,,
\eq
\beq{1.26}
(\bp+\bL)\Psi=0\,,
\eq
\beq{1.27}
(\p_j+M_j)\Psi=0\,.
\eq

\subsection{  Conservation laws II}.

The matrix equation (\ref{1.25}) allows to write down the conservation laws.
Its generic solutions can be represented in the form
\beq{1.28}
\Psi(x)=(I+R)\exp\left(-\f1{k}\int_0^x{\mathcal S} dx'\right)\,,
\eq
where $I$ is the identity matrix, $R$ is an off-diagonal periodic matrix and
${\mathcal S}=\di({\mathcal S}^1,\ldots,{\mathcal S}^N)$ is a diagonal matrix.
The equation (\ref{1.25}) means that $L$ can be gauge transformed to the
diagonal form
\beq{1.28a}
(I+R){\mathcal S}=k\p_x(I+R)+L(I+R).
\eq
Consider this equation in neighborhood of a point on $\Si_n$ with a local
coordinate $z$. Assume, for simplicity, that it is a pole of the Lax operator
and $k$ is a constant. In particular, it follows from (\ref{1.11})
 that $r_\al^0=0$ and the coadjoint orbits have the form
${\cal O}_\al=\{p_\al=-gp_\al^0g^{-1}\}$.
Then substitute into (\ref{1.28a}) the series expansions
$$
L(z)=L_{-1}z^{-1}+L_0+L_1z+\ldots, ~~(L_{-1}={\rm res}L=p),
$$
$$
{\mathcal S}(z)={\mathcal S}_{-1}z^{-1}+{\mathcal S}_0+{\mathcal S}_1z+\ldots,
$$
$$
(I+R)(z)=h+R_1z+R_2z^2+\ldots,~~(\di(R_m)\equiv 0).
$$
It follows from (\ref{1.28}) that the diagonal matrix elements ${\mathcal S}_j^m$ are
the densities of the conservation laws
$$
\log H_{j,l}=\oint  {\mathcal S}_j^ldx.
$$

We present a recurrence
procedure to define the diagonal matrices ${\mathcal S}_j$.
On the first step we find that
\beq{1.29}
{\mathcal S}_{-1}=h^{-1}L_{-1}h=h^{-1}ph,~~p=L_{-1}={\rm Res}~L_{z=0}\,.
\eq
In other words the diagonal matrix $S_{-1}$ determines the orbit located
at the point $z=0$.

In the general case we get the following equation
$$
{\mathcal S}_k+[h^{-1}R_{k+1},{\mathcal S}_{-1}]=
h^{-1}k\p_xR_k+h^{-1}\sum_{l=1}^kL_{l-1}R_{k-l+l}-R_l{\mathcal S}_{k-l}
h^{-1}L_kh\,.
$$
Separating the diagonal and the off-diagonal parts allows us to express
${\mathcal S}_k$ and $R_k$ in terms of the lower coefficients
\beq{1.30}
{\mathcal S}_k=\left(h^{-1}k\p_xR_k+
h^{-1}\sum_{l=0}^{k-1}L_lR_{k-l}+h^{-1}\sum_{l=1}^kL_{l-1}R_{k-l+l}-R_l{\mathcal S}_{k-l}
h^{-1}L_kh\right)_{diag}\,,
\eq
\beq{1.31}
[h^{-1}R_{k},{\mathcal S}_{-1}]=
\left(h^{-1}k\p_xR_{k-1}+h^{-1}\sum_{l=1}^kL_{l-1}R_{k-l+l}-R_l{\mathcal S}_{k-l}
h^{-1}L_kh\right)_{nondiag}\,.
\eq
In particular,
\beq{1.32}
{\mathcal S}_0=(h^{-1}k\p_xh+h^{-1}L_0h)_{diag}\,,
\eq
\beq{1.33}
{\mathcal S}_1=\left(h^{-1}k\p_xR_1+h^{-1}L_1h-h^{-1}R_1{\mathcal S}_0+h^{-1}L_0R_1\right)_{diag}\,,
\eq
where $R_1$ is defined by the equation
\beq{1.34}
[h^{-1}R_1,{\mathcal S}_{-1}] =(h^{-1}k\p_xh+h^{-1}L_0h)_{nondiag}\,.
\eq

\subsection{ Hamiltonians in $\SL$ case.}

Let us perform the gauge transformation
\beq{b1}
f^{-1}L f+kf^{-1}\p_x f=L'\,,
\eq
with $f$ defined as follows:
\beq{b2}
f=
\left(
\begin{array}{l}
\ \ \ \sqrt{L_{12}}\ \ \ \ \ \ \ \ \ \ \ \ \ \ \ \ \ \ \
 \ \ \ \ \ \ \ \ 0
\\
-\frac{L_{11}}{\sqrt{L_{12}}}-k\frac{\p_x \sqrt{L_{12}}}
{L_{12}}\ \ \ \ \ \ \ \frac{1}{\sqrt{L_{12}}}
\end{array}
\right)\,.
\eq
Then the Lax matrix $L$ is transformed into
\beq{b3}
L'=
\left(
\begin{array}{l}
\ 0 \ \ \ \  1
\\
\ T \ \ \ \ 0
\end{array}
\right)\,,
\eq
where
\beq{b4}
T=L_{21}L_{12}+L_{11}^2+k\frac{L_{11}\p_xL_{12}}
{L_{12}}-k\p_xL_{11}-\frac{1}{2}k^2
\frac{\p_x^2L_{12}}{L_{12}}+\frac{3}{4}k^2
\frac{(\p_x{L_{12}})^2}{L_{12}^2}\,.
\eq
The linear problem
\beq{b5.1}
\left\{
\begin{array}{l}
(xk\p_x+L')\psi=0\,,
\\
(\p_j+M_j')\psi=0\,,
\end{array}
\right.
\eq
where $\psi$ is the Bloch wave function $\psi=exp\{-i\oint\chi\}$,
leads to the Riccati equation:
\beq{b7}
ik\p_x\chi-\chi^2+T=0.
\eq
The decomposition of $\chi(z)$ provides densities of the conservation laws
(see \cite{DMN}):
\beq{b8}
\chi=\sum\limits_{k=-1}^{\infty}z^k\chi_k\,,
\eq
\beq{b9}
H_k\sim\oint dx\chi_{k-1}\,.
\eq
The values of $\chi_k$ can be found from (\ref{b7}) using the expression
(\ref{b4}) for
 $T(z)=\sum\limits_{k=-2}^{\infty}z^k T_{k}$ in a neighborhood of zero.
For $k=-2,\ -1$ and $0$ we have:
\beq{b10}
\left\{
\begin{array}{l}
\chi_{-1}=\sqrt{T_{-2}}=\sqrt{h}\,,
\\
2\sqrt{h}\chi_{0}=T_{-1}+ik\p_x\chi_{-1}=T_{-1}\,,
\\
2\sqrt{h}\chi_{1}=T_0+ik\p_x\chi-\chi_0^2\,.
\end{array}
\right.
\eq
In Subsections 7.2, 7.3 below, explicit formulae for $T_k$ are used for
the computation
of the Hamiltonians for the elliptic 2d Calogero-Moser and the elliptic
 Gaudin models.

\section{$\hat{L}(\SLN) $-bundles over  elliptic curves with marked points}
\setcounter{equation}{0}

\subsection{General case.}

We apply the general construction to the $\hat{L}(\SLN) $-bundle over
elliptic curve $E_\tau$
with marked points $w_\al,~\al=1,\ldots,n$. It is a two-dimensional
generalization of the elliptic Gaudin model \cite{Ne}.
In particular, for one marked point $z=0$ we come to the
$N$-body elliptic CM field theory.

Let us construct
solutions of the moment equations (\ref{1.9}), taking for simplicity
at the marked points the orbits with vanishing central charges
$$
{\mathcal O}_\al=\left\{
p_{ij}^\al,~~
r_\al=0
\right\}\,.
$$
For elliptic curves one can fix the central charge as $k=1$.
For the stable bundles the gauge
transformation (\ref{1.2o}) allows to diagonalize $\bA$:
\beq{diag}
\bar{A}_{ij}=\delta_{ij}\frac{2\pi\sqrt{-1}}{\tau-\bar{\tau}}u_i \,.
\eq
Then the Lax operator $L^G$ should satisfy (\ref{1.14}). It takes the form:
\beq{L}
\begin{array}{c}
L^G_{ij}=-\frac{\delta_{ij}}{2\pi\sqrt{-1}}\left(
\frac{v_i}{2}+\sum\limits_{\al}p^\al_{ii}\left(2\pi\sqrt{-1}
\frac{z-\bar{z}}{\tau-\bar{\tau}}+E_1(z-w_\al)\right)\right)-
\\
-\frac{1-\de_{ij}}{2\pi\sqrt{-1}}\sum\limits_{\al}p^\al_{ij}
\bfe\left(\frac{z-w_\al-(\bar{z}-\bar{w}_\al)}{\tau-\bar{\tau}}
u_{ij}\right)\phi(u_{ij},z-w_\al),~~(u_{ij}=u_i-u_j)\,.
\end{array}
\eq
By the quasiperiodic gauge transform
\beq{rt}
f=\hbox{diag}\{\bfe(\frac{z-\bar{z}}{\tau-\bar{\tau}}u_i)\}
\eq
one comes to
the holomorphic quasiperiodic  Lax operator
\beq{holL}
l^G_{ij}(z)=-\frac{\delta_{ij}}{2\pi\sqrt{-1}}\left(
\frac{v_i}{2}+\sum\limits_{\al}p^\al_{ii}E_1(z-w_\al)\right)-
\frac{1-\de_{ij}}{2\pi\sqrt{-1}}\sum\limits_{\al}p^\al_{ij}
\phi(u_{ij},z-w_\al)\,.
\eq
reducing the moment map equation to the diagonal gives the additional
constraint
\beq{const}
\frac{1}{2\pi\sqrt{-1}}\sum\limits_{\al}p_{ii}^\al=\p_x u_i\,.
\eq

\subsection{$\hat{L}(\SL) $-bundles over  elliptic curves with marked points.}

In this subsection we study 2-body elliptic Calogero field theory
in details.

\paragraph {The operator $L$.}
 According to (\ref{holL}) the holomorphic Lax operator is
\beq{5.5}
\left\{
\begin{array}{l}
l_{11}^G=-\frac{v}{4\pi \sqrt{-1}}-\sum\limits_\al
\frac{p_{11}^\al}{2\pi \sqrt{-1}}E_1(z-w_\al)\,,
\\
l_{12}^G=-\sum\limits_{\al}\frac{p_{12}^\al}{2\pi\sqrt{-1}}\phi(2u,z-w_\al)\,,
\\
l_{21}^G=-\sum\limits_{\al}\frac{p_{21}^\al}{2\pi\sqrt{-1}}\phi(-2u,z-w_\al)\,,
\end{array}
\right.
\eq
with the additional constraint (\ref{const})
\beq{5.3}
\frac{1}{2\pi\sqrt{-1}}\sum\limits_\al p_{11}^\al=u_x\,.
\eq

We still have the freedom to fix the gauge with respect to the action of the
diagonal subgroup. The corresponding moment map is (\ref{5.3}).

For the one marked point  $w_1=0$ the corresponding orbit is
\beq{5.6}
p=2\pi\sqrt{-1}
\left(
\begin{array}{l}
\ u_x\ \ \ -\nu
\\
-\nu\ \ \ -u_x
\end{array}
\right),
\eq
where $\nu=$const. is the result of the gauge fixing. In this case the
Lax operator is a
 $2$d generalization of the Lax operator for the two-body CM model:
\beq{5.7}
L^{CM}_{2D}=
\left(
\begin{array}{l}
-\frac{1}{4\pi\sqrt{-1}}v-u_xE_{1}(z)\ \ \ \ \ \ \ {\nu}\phi(2u,z)
\\
\ \ \
{\nu}\phi(-2u,z)\ \ \ \ \ \ \ \frac{1}{4\pi\sqrt{-1}}v+u_xE_{1}(z)
\end{array}
\right)\,,
\eq
This operator is still periodic under the shift $z\to z+1$ and
$$
L^{CM}_{2D}(z+\tau)={\bf e}(u)L^{CM}_{2D}(z){\bf e}(-u)
+{\bf e}(u)\p_x{\bf e}(-u)\,,
$$
where ${\bf e}(u)=\di(\exp u,\exp -u)$.

\paragraph{ Hamiltonians for the 2d elliptic ${\rm sl}(2,{\mathbb C})$
CM model.}
In this case the coefficients $T_k$ are (see (\ref{b4})-(\ref{b10})):
\beq{b6}
\left\{
\begin{array}{l}
T_{-2}^{CM}=u_x^2+\nu^2=h
\\
T_{-1}^{CM}=2\frac{v}{4\pi\sqrt{-1}}u_x-\frac{\nu_x}{\nu}u_x+u_{xx}
\\
T_{0}^{CM}=-\frac{v^2}{16\pi^2}+(2u_x^2-\nu^2)\wp(2u)-\frac{v}{4\pi\sqrt{-1}}
\frac{\nu_x}{\nu}+\frac{1}{4}(\frac{\nu_x}{\nu})^2
\end{array}
\right.\,,
\eq
where $h$ is the Casimir function, fixing the coadjoint orbit at the
marked point. It can be  chosen as a constant. Thus, we have
$$
\nu^2=h-u_x^2.
$$

The next order Hamiltonian is quadratic
\beq{b6.1}
H_{-1}^{CM}=
\oint\frac{v}{2\pi\sqrt{-1}}u_x-\frac{\nu_x}{\nu}u_x\,.
\eq
It can be written in the following way:
\beq{c3}
H_{-1}^{CM}=\oint\frac{v}{2\pi\sqrt{-1}}u_x+\frac{u_{xx}h}{\nu^2}\,.
\eq
Since $\{\oint dx\frac{u_{xx}}{\nu^2},v(y)\}=0$,
the equations of motion are:
\beq{c4}
\left\{
\begin{array}{l}
u_t=\frac{1}{2\pi\sqrt{-1}}u_x\,,
\\
v_t=\frac{1}{2\pi\sqrt{-1}}v_x\,.
\end{array}
\right.
\eq
Note that the L-M pair is simple in this case: $M=\frac{1}{2\pi\sqrt{-1}}L$.

The first nontrivial Hamiltonian $H_0$ is quadratic in the momenta field $v$.
It is a two-dimensional generalization of the quadratic CM Hamiltonian
\beq{b11}
H_0^{CM}=\oint dx2\sqrt{h}\chi_{1}=\oint dx
(T_0-\frac{1}{4h}T_{-1}^2)\,.
\eq
A direct evaluation yields:
\beq{b12}
T_0^{CM}-\frac{1}{4h}(T_{-1}^{CM})^2=-\frac{v^2}{16\pi^2}(1-\frac{u_x^2}{h})
+(3u_x^2-h)\wp(2u)-\frac{u_{xx}^2}{4\nu^2}\,.
\eq

The equations of motion produced by $H_0^{CM}$ are:
\beq{b13}
u_t=-\frac{v}{8\pi^2}(1-\frac{u_x^2}{h})\,,
\eq
$$
v_t=\frac{1}{8\pi^2 h}\p_x(v^2u_x)-2(3u_x^2-h)\wp'(2u)+
6\p_x(u_x\wp(2u))+
\frac{1}{2}\p_x(\frac{u_{xxx}\nu-\nu_xu_{xx}}{\nu^3})\,.
$$
It is reduced to the two-body elliptic CM system for the $x$-independent
fields.

\paragraph{The L-M pair for the 2d elliptic ${\rm sl}(2,{\mathbb C})$
 CM model.}
The equations of motion (\ref{b13}) produced by the quadratic Hamiltonian
$H_0^{CM}$ can be represented in a form of the Zakharov-Shabat
equation with the $L$ matrix defined by (\ref{5.7}) and the $M$
matrix given as follows:
\beq{c2}
\left\{
\begin{array}{l}
M_{11}=-u_tE_1(z)-\frac{1}{4\pi\sqrt{-1}}\left(
\frac{1}{8\pi^2h}v^2u_x+6u_x\wp(2u)+
\frac{u_{xxx}\nu-\nu_xu_{xx}}{2\nu^3}\right)+\\+
\frac{u_x}{2\pi\sqrt{-1}}(E_2(2u)-E_2(z))\,,
\\
M_{12}=-\frac{\nu}{2\pi\sqrt{-1}}\phi'(2u,z)+
\left(\frac{\nu}{2\pi\sqrt{-1}}E_1(z)
+\frac{vu_x\nu}{8\pi^2h}
-\frac{1}{4\pi\sqrt{-1}}\frac{u_{xx}}{\nu}\right)\phi(2u,z)\,,
\\
M_{21}=-\frac{\nu}{2\pi\sqrt{-1}}\phi'(-2u,z)+
\left(\frac{\nu}{2\pi\sqrt{-1}}E_1(z)+
\frac{vu_x\nu}{8\pi^2h}
+\frac{1}{4\pi\sqrt{-1}}\frac{u_{xx}}{\nu}\right)\phi(-2u,z)\,.
\end{array}
\right.
\eq
See Appendix C for details of the proof.
This construction completes the description of right vertical arrow in
Fig.1

\paragraph {2d CM - LL correspondence}.
The upper modification that produces the map of the elliptic CM system into the
elliptic rotator (\ref{8.4}), (\ref{8.5})
 works in the two-dimensional case as well.

The two-dimensional extension of the $\SL$-elliptic rotator is
 the Landau-Lifshitz (LL) equation
\beq{LL}
\p_t\bfS=\frac{1}{2}[\bfS,J(\bfS)]+\oh[\bfS,\p_{xx}\bfS].
\eq
This equation can be fitted in the Zakharov-Shabat form
\cite{Sk}. The Lax operator $L^{LL}$ has the same form
as for the $\SL$ elliptic rotator $L^{rot}$ (\ref{8.7}).
 For ${{\rm sl}(2,{\mathbb C})}$
 the basis of the sigma matrices coincides with the basis
of the sin-algebra and $L^{LL}$ takes the form
$$
L=\sum_au_a(z)S_a\si_a,
$$
$$
u_1=\varphi\left[\begin{array}{c}0\\1\end{array}\right](z)\,,~
u_2=\varphi\left[\begin{array}{c}1\\1\end{array}\right](z)\,,~
u_3=\varphi\left[\begin{array}{c}1\\0\end{array}\right](z)\,.
$$
The $M^{LL}$ operator has a very simple extension
$$
M^{LL}=M^{rot}-L^{rot}E_1(z)+\sum_au_a(z)\tr(\si_a[\bfS,\p_x\bfS])\si_a\,.
$$
It is easy to check that the Zakharov-Shabat equation leads to
(\ref{LL}) if
$$
\sum_a S_a^2=1\,.
$$
 Thereby be have defined the right vertical arrow in Fig.1.

Consider the upper modification $\Xi_{2D}$ that has the same
quasi-periodicity as $\Xi$ but corresponds to the residue $p$
(\ref{5.6}) of $L_{2D}^{CM}$ (\ref{5.7}).
Then the Lax operator for the LL system is the result of the upper
modification
\beq{UM}
L^{LL}=\Xi_{2D}\p_x\Xi_{2D}^{-1}+\Xi_{2D} L_{2D}^{CM}\Xi_{2D}^{-1}\,.
\eq
 It means that we can
pass from the $CM$ fields $v(x,t),u(x,t)$ and the constant $\nu$ to
the LL fields $\bfS=(S_1,S_2,S_3)$ with the orbit fixing condition
$$
\sum_a S_a^2=-\f1{2\pi^2}(u_x^2+\nu^2)=1\,.
$$
It completes the description of the diagram on Fig.1.

\paragraph{ Relations with the Sinh-G equation and the nonlinear Schr\"{o}dinger
equation.}
It is known that the LL model is universal; it contains as a special limit
the Sinh-Gordon and the Nonlinear Schr\"{o}dinger models \cite{FT2}.
In this way they can be derived within the 2d CM system.

The scaling limit in the CM model is a combination of the trigonometric limit
$Im\tau\rightarrow\infty$ with shifts of coordinates: $u=U+\frac{1}{2}
Im\tau$ and renormalization of
the coupling constant $\nu=\bar{\nu} e^{\frac{1}{2}Im\tau}$ \cite{I}.
This procedure applied to the 2d elliptic CM Hamiltonian
yields the sinh-Gordon system:
\beq{d1}
H_{SG}=-\frac{v^2}{16\pi^2}-\bar{\nu}^2(e^{2U}+e^{-2U})+\frac{U_x^2}{4}\,.
\eq
The equations of motion are:
\beq{d2}
\left\{
\begin{array}{l}
U_t=-\frac{v}{8\pi^2}\,,
\\
v_t=2\bar{\nu}^2(e^{2U}-e^{-2U})+\frac{1}{2}U_{xx}\,.
\end{array}
\right.
\eq
The L-M pair is:
\beq{d3}
L^{SG}=
\left(
\begin{array}{l}
-\frac{v}{4\pi\sqrt{-1}}-\frac{1}{2}U_x\ \ \ \ \ \bar{\nu}(1-e^{2U}Z)
\\
\bar{\nu}(\frac{1}{Z}-e^{-2U}) \ \ \ \ \ \ \frac{v}{4\pi\sqrt{-1}}+
\frac{1}{2}U_x
\end{array}
\right)\,.
\eq
\beq{d4}
M^{SG}=
\left(
\begin{array}{l}
-\frac{U_t}{2}-\frac{1}{8\pi\sqrt{-1}}U_x\ \ \ \ \
\frac{\bar{\nu}}{4\pi\sqrt{-1}}
(1+e^{2U}Z)
\\
\frac{\bar{\nu}}{4\pi\sqrt{-1}}(e^{-2U}+\frac{1}{Z})\ \ \ \ \ \
\frac{U_t}{2}+\frac{1}{8\pi\sqrt{-1}}U_x
\end{array}
\right)\,.
\eq

Let us consider $2d$ CM theory for $N=2$ in the rational limit when the both periods
of the basic spectral curve go to infinity.  The upper modification
(\ref{UM}) transforms this system in the Heisenberg magnetic. Then using the
non-singular gauge transform from Ref.\,\cite{FT2} we come to the nonlinear Schr\"{o}dinger
equation.

\subsection{ Hamiltonians for the 2d elliptic Gaudin model.}
Using (\ref{ad3}) we obtain the Hamiltonian:
\beq{e5}
H_{-1,a}^G=2\frac{v}{4\pi\sqrt{-1}}\frac{p_{11}^a}{2\pi\sqrt{-1}}+
2\sum\limits_b
\frac{p_{11}^a}{2\pi\sqrt{-1}}\frac{p_{11}^b}{2\pi\sqrt{-1}}E_1(z_a-z_b)-
\eq
$$
-\sum\limits_{a\neq b}\frac{p_{12}^ap_{21}^b}{(2\pi\sqrt{-1})^2}
\phi(2u,z_b-z_a)
+\sum\limits_{a\neq b}\frac{p_{12}^bp_{21}^a}{(2\pi\sqrt{-1})^2}
\phi(2u,z_a-z_b)
-\frac{p_{11}^a}{2\pi\sqrt{-1}}\frac{\p_x p_{12}^a}{p_{12}}\,.
$$
The last term makes the above Hamiltonian different from the
one-dimensional version.

Let us consider the ${\rm sl}(2, {\mathbb C})$
case with two marked points on the elliptic curve.

We will use the following notations:
\beq{e6}
\left\{
\begin{array}{l}
p_{11}^1=2\pi\sqrt{-1}\gamma_1,\ \ \ p_{11}^2=2\pi\sqrt{-1}\gamma_2\,,
\\
p_{12}^1=-2\pi\sqrt{-1}\nu_+,\ \ \ p_{21}^1=-2\pi\sqrt{-1}\nu_-\,,
\\
p_{12}^2=-2\pi\sqrt{-1}\mu_+,\ \ \ p_{21}^2=-2\pi\sqrt{-1}\mu_-\,.
\end{array}
\right.
\eq

The $L$ matrix is:
\beq{e7}
\left\{
\begin{array}{l}
l_{11}^G=-\frac{v}{4\pi\sqrt{-1}}-\gamma_1E_1(z-z_1)-\gamma_2E_1(z-z_2)\,,
\\
l_{12}^G=\nu\phi(2u,z-z_1)+\mu_+\phi(2u,z-z_2)\,,
\\
l_{21}^G=\nu\phi(-2u,z-z_1)+\mu_-\phi(-2u,z-z_2)\,.
\end{array}
\right.
\eq
The solution exists if
\beq{e8}
\gamma_1+\gamma_2=u_x \,.
\eq

The gauge fixing condition is chosen to be
\beq{e9}
\nu_+=\nu_-=\nu.
\eq
We fix the Casimir elements $h_1=\gamma_1^2+\nu^2$ and $h_2=\gamma_2^2+
\mu_+\mu_-$ to be constants:$\ h_1,\ h_2\in {\mathbb C}$.

On the reduced phase space there are two independent fields besides
$u$ and $v$. Let them be for example $\nu$ and $\mu_+$, then
\beq{e9.1}
\left\{
\begin{array}{l}
\gamma_1=\sqrt{h_1-\nu^2}\,,
\\
\gamma_2=u_x-\sqrt{h_1-\nu^2}\,,
\\
\mu_-=\frac{1}{\mu_+}(h_2-(u_x-\sqrt{h_1-\nu^2})^2)\,.
\end{array}
\right.
\eq
However we are going to use all kinds of variables in order to make
the formulae more transparent.
The non-trivial brackets on the reduced phase space are:
\beq{e12}
\begin{array}{c}
\{v(x),u(y)\}=\delta(x-y)\,,\ \ \{v(x),\gamma_1(y)\}=-\delta'(x-y)\,,\ \
\{v(x),\nu(y)\}=\frac{\gamma_1}{\nu}\delta'(x-y)\,,
\\
\{\mu_+(x),\gamma_1(y)\}=-\frac{1}{2\pi\sqrt{-1}}\mu_+\delta(x-y)\,,
\ \ \{\mu_+(x),\mu_-(y)\}=-2\frac{1}{2\pi\sqrt{-1}}\gamma_2\delta(x-y)\,,
\\
\{\mu_+(x),\gamma_2(y)\}=\frac{1}{2\pi\sqrt{-1}}\mu_+\delta(x-y)\,,\ \ \ \
\{\mu_+(x),\nu(y)\}=\frac{1}{2\pi\sqrt{-1}}
\frac{\gamma_1}{\nu}\mu_+\delta(x-y)\,,
\\
\{\nu(x),\mu_-(y)\}=\frac{1}{2\pi\sqrt{-1}}
\frac{\gamma_1}{\nu}\mu_-\delta(x-y)\,.
\end{array}
\eq
The Hamiltonian is:
\beq{e14}
\begin{array}{c}
H_{-1}^G=\oint dx\left(2\gamma_1\frac{v}{4\pi\sqrt{-1}}-
\gamma_1\frac{\nu_x}{\nu}+\p_x\gamma_1
+\nu\mu_+\phi(2u,z_1-z_2)-\nu\mu_-\phi(2u,z_2-z_1)+\right.
\\
\left.
-2\gamma_1\gamma_2E_1(z_1-z_2)\right)\,.
\end{array}
\eq
The equations of motion are:
\beq{e15}
\left\{
\begin{array}{l}
\p_t u(x)=\frac{1}{2\pi\sqrt{-1}}\gamma_1(x)\,,
\\
\p_t v(x)=\frac{1}{2\pi\sqrt{-1}}v_x-
\p_x(\frac{\gamma_1\mu_+}{\nu}\phi(2u,z_1-z_2))+
\p_x(\frac{\gamma_1\mu_-}{\nu}\phi(2u,z_2-z_1))-
\\
-2\nu\mu_+\phi'(2u,z_1-z_2)+2\nu\mu_-\phi'(2u,z_2-z_1)-
\p_x\left(\frac{\gamma_1\p_x\gamma_1}{\nu^2}\right)\,,
\\
\p_t \nu=-\frac{1}{2\pi\sqrt{-1}}\p_x\left(\frac{\gamma_1^2}{\nu}\right)
+\frac{\gamma_1}{2\pi\sqrt{-1}\nu}(\mu_+\phi(2u,z_1-z_2)-\mu_-
\phi(2u,z_2-z_1))\,,
\\
\p_t \mu_+=\frac{1}{2\pi\sqrt{-1}}(2\frac{v}{4\pi\sqrt{-1}}\mu_+
2\mu_+(\gamma_2-\gamma_1)E_1(z_1-z_2))-\frac{2\nu\gamma_2}{2\pi\sqrt{-1}}
\phi(2u,z_2-z_1)-
\\
-\frac{\gamma_1\mu_+}{2\pi\sqrt{-1}\nu}(\mu_+\phi(2u,z_1-z_2)
-\mu_-\phi(2u,z_2-z_1))\,.
\end{array}
\right.
\eq
The quadratic Hamiltonian is the direct generalization of (\ref{b12})
\beq{e16}
H_0^G=\oint dx2\sqrt{h_1}\chi_1=\oint dx
(T_0-\frac{1}{4h_1}T_{-1}^2)\,,
\eq
where
\beq{e17}
\begin{array}{c}
2\sqrt{h_1}\chi_1=
\\
-\frac{v^2}{16\pi^2}(1-\frac{\gamma_1^2}{h_1})
+(2u_x\gamma_1-\nu^2)\wp(2u)-
\frac{(\p_x\gamma_1)^2}{4\nu^2}+\mu_+\mu_-(E_2(z_1-z_2)-E_2(2u))+
\\
+4\eta_1\gamma_1\gamma_2

+\nu\mu_-\phi(2u,z_2-z_1)(E_1(z_1-z_2)-E_1(2u)+E_1(2u+z_2-z_1))-
\\
-\nu\mu_+\phi(2u,z_1-z_2)(E_1(z_1-z_2)+E_1(2u)-E_1(2u+z_1-z_2))+
\gamma_2^2E_1^2(z_1-z_2)+
\\
+2\gamma_2\frac{v}{4\pi\sqrt{-1}}E_1(z_1-z_2)-
\gamma_2\frac{\nu_x}{\nu}E_1(z_1-z_2)
+\gamma_1\frac{\mu_+}{\nu^2}\phi(2u,z_1-z_2)-
\\
-\gamma_1\phi(2u,z_1-z_2)[\frac{\p_x\mu_+}{\nu}+2u_x\frac{\mu_+}{\nu}
(E_1(z_1-z_2+2u)-E_1(2u))]-
\\
-\frac{1}{4h_1}(\nu\mu_+\phi(2u,z_1-z_2)-\nu\mu_+\phi(2u,z_2-z_1)
+2\gamma_1\gamma_2E_1(z_1-z_2))^2-
\\
-\frac{1}{2h_1}\left(\nu\mu_+\phi(2u,z_1-z_2)-\nu\mu_+\phi(2u,z_2-z_1)+
\right.
\\
\left.
+2\gamma_1\gamma_2E_1(z_1-z_2)\right)
(2\gamma_1\frac{v}{4\pi\sqrt{-1}}-\gamma_1\frac{\nu_x}{\nu}+\p_x\gamma_1)\,.
\end{array}
\eq

\section{Conclusion}
Here we briefly summarize the results of our analysis and discuss some
unsolved related problems.
The following two subjects were investigating in the paper.

({\bf i}) We have constructed symplectic maps between Hitchin systems related to holomorphic
bundles of different degrees.
It allowed us to construct the B\"{a}cklund transformations in the
Hitchin systems defined over Riemann curves with  marked points.
We applied the general scheme to the elliptic CM systems and constructed
the symplectic map to an integrable $\SLN$ Euler-Arnold top (the elliptic
$\SLN$-rotator).
The open problem is to write down the explicit expressions
for the spin variables in
 terms of the CM phase space for an arbitrary $N$ as it was done
for the case $N=2$ (\ref{S}).  It should help to construct the
B\"{a}cklund transformations for the CM systems explicitly, and more generally,
to construct the generating function for them.
The later can be considered as the integrable discrete time mapping
\cite{Ve}.

(ii) We have proposed a generalization of the Hitchin approach
to 2d integrable
theories related to holomorphic bundles of infinite rank. The main example
is the integrable two-dimensional version of the two-body elliptic CM system.
The upper modification allows to define the symplectic map to the
Landau-Lifshitz equation and to find, in principle,
 the B\"{a}cklund transformations in the field theories.\\
It will be extremely interesting to find the 2d generalization
of the $\SLN$-rotator for $N>2$ (the matrix LL equation).\\
There is another point of view
on the 2d generalizations of the Hitchin systems.
One can try to define them starting from holomorphic bundles
over complex surfaces, that are fibrations over Riemann curves.
In this case the spectral parameter lives on the base of the fibration,
while the space variable lives on the fibers. It will be interesting
to analyze, for example, the known solutions of the LL equation from this point
of view.

\section{Acknowledgments} We would like to thank K.Hasegawa and
I.Krichever for illuminating discussions and A.Shabat, T.Takebe,
V.Sokolov and A.Zabrodin for useful remarks and H.Gangl
for the careful reading the manuscript.
The remarks of the referee allow us to improve the text in essential way.
 We are grateful for the hospitality
of the Max-Planck-Institut f\"{u}r Mathematik (Bonn), where the paper was
partly prepared during the visits of A.L. and M.O.
The work of all authors
was supported by the grant 00-15-96557 of the scientific schools and
RFBR-01-01-00539 (A.L.), RFBR-00-02-16530 (M.O.,A.Z.), INTAS-00-00561 (A.Z.),
INTAS-99-01782 (M.O.).

\section{Appendix}
\subsection{Appendix A. Sin-Algebra.}
\setcounter{equation}{0}
\def\theequation{A.\arabic{equation}}

\beq{AA0}
{\bf e}(z)=\exp (2\pi \sqrt{-1}z)\,,
\eq
\beq{AA11}
Q=\di({\bf e}(1/N),\ldots,{\bf e}(m/N),\ldots,1)\,,
\eq
\beq{AA2}
\La=
\left(\begin{array}{ccccc}
0&1&0&\cdots&0\\
0&0&1&\cdots&0\\
\vdots&\vdots&\ddots&\ddots&\vdots\\
0&0&0&\cdots&1\\
1&0&0&\cdots&0
\end{array}\right)\,,
\eq
\beq{AA3}
E_{mn}=\bfe(\frac{mn}{2N})Q^m\La^n,~(m=0,\ldots,N-1,~n=0,\ldots,N-1,
({\rm mod} N)~
m^2+n^2\neq 0)
\eq
is the basis in $\sln$. The commutation relations in this basis take the form
\beq{AA4}
[E_{sk},E_{nj}]=2\sqrt{-1}\sin \frac{\pi}{N}(kn-sj)E_{s+n,k+j}\,,
\eq
\beq{AA5}
\tr (E_{sk}E_{nj})=\de_{s,-n}\de_{k,-j}N\,.
\eq

\subsection{Appendix B. Elliptic functions.}
\setcounter{equation}{0}
\def\theequation{B.\arabic{equation}}

We summarize the main formulae for elliptic functions, borrowed
mainly from \cite{We} and \cite{Mu}.
We assume that $q=\exp 2\pi i\tau$, where $\tau$ is the modular parameter
of the elliptic curve $E_\tau$.

The basic element is the theta  function:
\beq{A.1a}
\vth(z|\tau)=q^{\frac
{1}{8}}\sum_{n\in {\bf Z}}(-1)^ne^{\pi i(n(n+1)\tau+2nz)}=
\eq
$$
q^{\frac{1}{8}}e^{-\frac{i\pi}{4}} (e^{i\pi z}-e^{-i\pi z})
\prod_{n=1}^\infty(1-q^n)(1-q^ne^{2i\pi z})(1-q^ne^{-2i\pi z})\,.
 $$
\bigskip

{\sl The  Eisenstein functions}
\beq{A.1}
E_1(z|\tau)=\p_z\log\vth(z|\tau), ~~E_1(z|\tau)\sim\f1{z}-2\eta_1z,
\eq
where
\beq{A.6}
\eta_1(\tau)=\ze(\frac{1}{2})=
\eq
$$
\frac{3}{\pi^2}\sum_{m=-\infty}^{\infty}\sum_{n=-\infty}^{\infty '}
\frac{1}{(m\tau+n)^2}=\frac{24}{2\pi i}\frac{\eta'(\tau)}{\eta(\tau)}\,,
$$
where
$$
\eta(\tau)=q^{\frac{1}{24}}\prod_{n>0}(1-q^n)\,.
$$
is the Dedekind function.

\beq{A.2}
E_2(z|\tau)=-\p_zE_1(z|\tau)=
\p_z^2\log\vth(z|\tau),
~~E_2(z|\tau)\sim\f1{z^2}+2\eta_1\,.
\eq

The next important function is
\beq{A.3}
\phi(u,z)=
\frac
{\vth(u+z)\vth'(0)}
{\vth(u)\vth(z)}\,.
\eq
It has a pole at $z=0$ and
\beq{A.3a}
\phi(u,z)=\frac{1}{z}+E_1(u)+\frac{z}{2}(E_1^2(u)-\wp(u))+\ldots\,,
\eq
and
\beq{A3b}
\phi(u,z)^{-1}\p_u\phi(u,z)=E_1(u+z)-E_1(u)\,.
\eq
The following formula plays an important role in checking
of the zero curvature equation:
\beq{aa4}
\phi''(u,z)=\phi(u,z)(E_2(z)-E_1^2(z)+2E_1(z)(E_1(u+z)-E_1(u))+2E_2(u)-
6\eta_1)\,.
\eq
It follows from:
\beq{aa5}
(E_1(z)+E_1(u)-E_1(z+u))^2=E_2(u)+E_2(z)+E_2(u+z)-6\eta_1\,.
\eq
\bigskip

{\sl Relations to the Weierstrass functions}
\beq{A.4}
\ze(z|\tau)=E_1(z|\tau)+2\eta_1(\tau)z\,,
\eq
\beq{A.5}
\wp(z|\tau)=E_2(z|\tau)-2\eta_1(\tau)\,,
\eq
\beq{A.7}
\phi(u,z)=\exp(-2\eta_1uz)
\frac
{\si(u+z)}{\si(u)\si(z)}\,,
\eq
\beq{A.7a}
\phi(u,z)\phi(-u,z)=\wp(z)-\wp(u)=E_2(z)-E_2(u)\,.
\eq

\bigskip

{\sl Particular values}

\beq{A7b}
E_1(\oh)=0,~~E_1(\frac{\tau}{2})=E_1(\frac{1+\tau}{2})=
-\pi\sqrt{-1}\,.
\eq
\bigskip

{\sl Series representations}

\beq{A.8}
E_1(z|\tau)=-2\pi i(\frac{1}{2}+\sum_{n\neq 0}
\frac{e^{2\pi iz}}{1-q^n})=
\eq
$$
-2\pi i(\sum_{n<0}\frac{1}{1-q^ne^{2\pi iz}}+
\sum_{n\geq 0}\frac{q^ne^{2\pi iz}}{1-q^ne^{2\pi iz}}+\frac{1}{2}),,
$$
\beq{A.9}
E_2(z|\tau)
=-4\pi^2\sum_{n\in{\bf Z}}\frac{q^ne^{2\pi iz}}{(1-q^ne^{2\pi iz})^2}\,,
\eq
\beq{A.10}
\phi(u,z)=2\pi i\sum_{n\in{\bf Z}}\frac
{e^{-2\pi inz}}{1-q^ne^{-2\pi iu}}.
\eq
 \bigskip

{\sl Parity}

\beq{P.1}
\vth(-z)=-\vth(z)\,,
\eq
\beq{P.2}
E_1(-z)=-E_1(z)\,,
\eq
\beq{P.3}
E_2(-z)=E_2(z)\,,
\eq
\beq{P.4}
\phi(u,z)=\phi(z,u)=-\phi(-u,-z)\,,
\eq
\bigskip

{\sl Quasi-periodicity}

\beq{A.11}
\vth(z+1)=-\vth(z)\,,~~~\vth(z+\tau)=-q^{-\oh}e^{-2\pi iz}\vth(z)\,,
\eq
\beq{A.12}
E_1(z+1)=E_1(z)\,,~~~E_1(z+\tau)=E_1(z)-2\pi i\,,
\eq
\beq{A.13}
E_2(z+1)=E_2(z)\,,~~~E_2(z+\tau)=E_2(z)\,,
\eq
\beq{A.14}
\phi(u+1,z)=\phi(u,z)\,,~~~\phi(u+\tau,z)=e^{-2\pi iz}\phi(u,z)\,.
\eq

 \bigskip

{\sl Addition formula}

\beq{ad1}
\phi(u,z)\p_v\phi(v,z)-\phi(v,z)\p_u\phi(u,z)=(E_2(v)-E_2(u))\phi(u+v,z)\,,
\eq
or
\beq{ad2}
\phi(u,z)\p_v\phi(v,z)-\phi(v,z)\p_u\phi(u,z)=(\wp(v)-\wp(u))\phi(u+v,z)\,.
\eq
The proof of (\ref{ad1}) is based on (\ref{A.3a}),(\ref{P.4}),
and (\ref{A.14}).
In fact, $\phi(u,z)$ satisfies more general relation which follows from the
Fay three-section formula
\beq{ad3}
\phi(u_1,z_1)\phi(u_2,z_2)-\phi(u_1+u_2,z_1)\phi(u_2,z_2-z_1)-
\phi(u_1+u_2,z_2)\phi(u_1,z_1-z_2)=0\,.
\eq
A particular case of this formula is
\beq{ad4}
\phi(u_1,z)\phi(u_2,z)-\phi(u_1+u_2,z)(E_1(u_1)+E_1(u_2))+
\p_z\phi(u_1+u_2,z)=0\,.
\eq
 \bigskip

{\sl Integrals}

\beq{A.28}
\int_{E_\tau}E_1(z|\tau)dzd\bz=0 \,.
\eq
  \bigskip

{\sl Theta functions with characteristics.}\\
For $a, b \in \Bbb Q$ put~:
\beq{A.30}
\theta{\left[\begin{array}{c}
a\\
b
\end{array}
\right]}(z , \tau )
=\sum_{j\in \Bbb Z}
{\bf e}\left((j+a)^2\frac\tau2+(j+a)(z+b)\right)\,.
\eq
In particular, the  function $\vth$ (\ref{A.1a}) is the theta
function with a characteristic
\beq{A.29}
\vartheta(x,\tau)=\theta\left[
\begin{array}{c}
1/2\\
1/2
\end{array}\right](x,\tau)\,.
\eq
One has
\beq{A.31}
\theta{\left[\begin{array}{c}
a\\
b
\end{array}
\right]}(z+1 , \tau )={\bf e}(a)
\theta{\left[\begin{array}{c}
a\\
b
\end{array}
\right]}(z  , \tau )\,,
\eq
\beq{A.32}
\theta{\left[\begin{array}{c}
a\\
b
\end{array}
\right]}(z+a'\tau , \tau )
={\bf e}\left(-{a'}^2\frac\tau2 -a'(z+b)\right)
\theta{\left[\begin{array}{c}
a+a'\\
b
\end{array}
\right]}(z , \tau )\,,
\eq
\beq{A.33}
\theta{\left[\begin{array}{c}
a+j\\
b
\end{array}
\right]}(z , \tau )=
\theta{\left[\begin{array}{c}
a\\
b
\end{array}
\right]}(z , \tau )\,,\quad j\in \Bbb Z\,.
\eq
For the simplicity we denote $\theta\left[\begin{array}{l}
a/2\\b/2\end{array}\right]=\theta_{ab}$.

The following identities are useful for the upper modification procedure
in  ${{\rm sl}(2,{\mathbb C})}$ case:
\beq{kz1}
\begin{array}{c}
\vtd(x,\tau)\vtc(y,\tau)+\vtd(y,\tau)\vtc(x,\tau)=
2\vtd(x+y,2\tau)\vtd(x-y,2\tau)\,,
\\
\vtd(x,\tau)\vtc(y,\tau)-\vtd(y,\tau)\vtc(x,\tau)=
2\vth(x+y,2\tau)\vth(x-y,2\tau)\,,
\\
\vtc(x,\tau)\vtc(y,\tau)+\vtd(y,\tau)\vtd(x,\tau)=
2\vtc(x+y,2\tau)\vtc(x-y,2\tau)\,,
\\
\vtc(x,\tau)\vtc(y,\tau)-\vtd(y,\tau)\vtd(x,\tau)=
2\vtb(x+y,2\tau)\vtb(x-y,2\tau)\,.
\end{array}
\eq

\beq{kz2}
\begin{array}{c}
2\vth(x,2\tau)\vtd(y,2\tau)=\vth(\frac{x+y}{2},\tau)\vtb(\frac{x-y}{2},\tau)
+\vtb(\frac{x+y}{2},\tau)\vth(\frac{x-y}{2},\tau)\,,
\\
2\vtc(x,2\tau)\vtb(y,2\tau)=\vth(\frac{x+y}{2},\tau)\vth(\frac{x-y}{2},\tau)
+\vtb(\frac{x+y}{2},\tau)\vtb(\frac{x-y}{2},\tau)\,,
\\
2\vtc(x,2\tau)\vtc(y,2\tau)=\vtc(\frac{x+y}{2},\tau)\vtc(\frac{x-y}{2},\tau)
+\vtd(\frac{x+y}{2},\tau)\vtd(\frac{x-y}{2},\tau)\,,
\\
2\vtb(x,2\tau)\vtb(y,2\tau)=\vtc(\frac{x+y}{2},\tau)\vtc(\frac{x-y}{2},\tau)
-\vtd(\frac{x+y}{2},\tau)\vtd(\frac{x-y}{2},\tau)\,.
\end{array}
\eq

\subsection{Appendix C: 2d ${\rm sl}(2,{\mathbb C})$ Calogero L-M pair.}
\setcounter{equation}{0}
\def\theequation{C.\arabic{equation}}

The Zakharov-Shabat equations in ${{\rm sl}(2,{\mathbb C})}$ case are:
\beq{c1}
\left\{
\begin{array}{l}
11:\p_tL_{11}-\p_xM_{11}=M_{21}L_{12}-M_{12}L_{21} \,,
\\
12:\p_tL_{12}-\p_xM_{12}=2L_{11}M_{12}-2L_{12}M_{11}\,,
\\
21:\p_tL_{21}-\p_xM_{21}=2M_{11}L_{21}-2L_{11}M_{21}\,.
\end{array}
\right.
\eq
Let the non diagonal terms in the M matrix be of the form:
\beq{aa1}
\left\{
\begin{array}{l}
M_{12}=c(x)\Phi'(2u,z)+(f_{12}^1(x)E_1(z)+f_{12}^0(x))\Phi(2u,z)\,,
\\
M_{21}=c(x)\Phi'(-2u,z)+(f_{21}^1(x)E_1(z)+f_{21}^0(x))\Phi(-2u,z)\,.
\end{array}
\right.
\eq
Then from the diagonal part of (\ref{c1}) we conclude:
\beq{aa2}
M_{11}=-u_tE_1(z)+\alpha(x)+\triangle M_{11} \,,
\eq
where
\beq{aa3}
\alpha(x)=-\frac{1}{4\pi\sqrt{-1}}\left(
\frac{1}{8\pi^2h}v^2u_x+6u_x\wp(2u)+
\frac{u_{xxx}\nu-\nu_xu_{xx}}{2\nu^3}\right)\,,
\eq
and
$\triangle M_{11}$ will be defined in the following. It is supposed
to be dependent on $E_2(2u)$ in order to cancel terms proportional to
$E_2(2u)$ and $E_2'(2u)$ in (\ref{c1}).

Using formula (\ref{aa4}) in the non diagonal part of (\ref{c1}), we arrive
to some conditions equivalent to cancellations of the terms
proportional to functions $\xi(2u,z)=E_1(2u+z)-E(2u)$,
$E_1(z)$, $E_1^2(z)$, $E_1(z)\xi(-2u,z)$:
\beq{aa6}
(12)
\ \ \ \
\left\{
\begin{array}{l}
E_1(z)\xi(2u,z):\ \ f_{12}^1=-c\,,
\\
E_1^2(z):\ \ f_{12}^1=-c\,,
\\
\xi(2u,z):\ \ 2\nu u_t-c_x-2u_xf_{12}^0=-2c\frac{v}{4\pi\sqrt{-1}}\,,
\\
E_1(z):\ \ -\p_x f_{12}^1=-2\frac{v}{4\pi\sqrt{-1}}f_{12}^1-2u_xf_{12}^0
+2u_t\nu    \,.
\end{array}
\right.
\eq
\beq{aa7}
(21)
\ \ \ \
\left\{
\begin{array}{l}
E_1(z)\xi(-2u,z):\ \ f_{12}^1=-c \,,
\\
E_1^2(z):\ \ f_{12}^1=-c \,,
\\
\xi(-2u,z):\ \ -2\nu u_t-c_x+2u_xf_{21}^0=2c\frac{v}{4\pi\sqrt{-1}}\,,
\\
E_1(z):\ \ -\p_x f_{21}^1=2\frac{v}{4\pi\sqrt{-1}}f_{12}^1+2u_xf_{21}^0
-2u_t\nu \,.
\end{array}
\right.
\eq
Thus
\beq{aa8}
\left\{
\begin{array}{l}
f_{12}^1=f_{21}^1=-c\,,
\\
2u_xf_{12}^0=2\nu u_t-c_x+2c\frac{v}{4\pi\sqrt{-1}}\,,
\\
2u_xf_{21}^0=2\nu u_t+c_x+2c\frac{v}{4\pi\sqrt{-1}}\,.
\end{array}
\right.
\eq
\beq{aa9}
\left\{
\begin{array}{l}
f_+=f_{21}^0+f_{12}^0=\frac{2}{u_x}(\nu u_t+c\frac{v}{4\pi\sqrt{-1}})\,,
\\
f_-=f_{21}^0-f_{12}^0=\frac{c_x}{u_x}\,.
\end{array}
\right.
\eq
The remaining parts of the non diagonal equations are:
\beq{aa10}
\left\{
\begin{array}{l}
\nu_t+12cu_x\eta_1-\p_xf_{12}^0-2cu_xE_2(z)-4cu_xE_2(2u)=\\=
-2\frac{v}{4\pi\sqrt{-1}}f_{12}^0-2\nu\alpha-2\nu\triangle M_{11}\,,
\\
-\nu_t+12cu_x\eta_1+\p_xf_{21}^0-2cu_xE_2(z)-4cu_xE_2(2u)=\\=
-2\frac{v}{4\pi\sqrt{-1}}f_{21}^0-2\nu\alpha-2\nu\triangle M_{11}\,.
\end{array}
\right.
\eq
Subtracting the above equations we have:
\beq{aa11}
2\frac{u_x}{\nu}\p_xu_t+\p_x f_+=-2\frac{v}{4\pi\sqrt{-1}}f_- \,.
\eq
Substituting $f_+$ and $f_-$ from (\ref{aa9}) into (\ref{aa11})
we arrive to the equation for $c$:
\beq{aa12}
\frac{u_x}{\nu}\p_xu_t+
\p_x(\frac{1}{u_x}(\nu u_t+c\frac{v}{4\pi\sqrt{-1}}))
=-\frac{v}{4\pi\sqrt{-1}}\frac{c_x}{u_x} \,.
\eq
Now some concrete equations of motion should be used.
For $H_{-1}^{CM}$ this equation yields $c\sim\sqrt{\frac{u_x}{v}}$.
However the coefficient of the proportionality appear to
be equal zero. For $H_0^{CM}$ (\ref{b12}) we have
$c=-\frac{\nu}{2\pi\sqrt{-1}}$.
\bigskip

\small{

}
\end{document}